\documentclass[a4paper,oneside]{amsart}

\pdfoutput=1
\usepackage{geometry}            
\usepackage[utf8]{inputenc} 
\usepackage[english]{babel} 
\usepackage{graphicx}
\usepackage{amsmath, mathtools}
\usepackage{amsthm}
\usepackage{enumerate}
\usepackage[textsize=tiny]{todonotes}
\usepackage[many]{tcolorbox}
\usepackage{tensor} 
\usepackage{here} 
\usepackage{subcaption}

\definecolor{light-gray}{gray}{0.85}
\definecolor{nicegreen}{RGB}{60,183,82}
\usepackage{amssymb}
\usepackage[breaklinks]{hyperref}
\definecolor{darkblue}{RGB}{0,0,127} 
\definecolor{darkgreen}{RGB}{0,180,0}
\definecolor{darkred}{RGB}{180,0,0}
\hypersetup{
	colorlinks, 
	linkcolor=darkblue, 
	citecolor=darkgreen, 
	filecolor=red, 
	urlcolor=blue,
	pdftitle={Gauging defects in quantum spin systems}, 
	pdfauthor={Jacob C. Bridgeman, Alexander Hahn, Tobias J. Osborne, and Ramona Wolf}
}

\allowdisplaybreaks

\usepackage{tikz}
\usetikzlibrary{decorations.pathreplacing,shapes.misc}
\usetikzlibrary{arrows,shapes,positioning}
\usetikzlibrary{decorations.markings}
\usetikzlibrary{through,calc}
\usetikzlibrary{patterns}
\usetikzlibrary{plotmarks}
\usetikzlibrary{automata}
\tikzstyle arrowstyle=[scale=1]
\usepackage{extarrows}

\definecolor{spacecadet}{HTML}{0D284C}
\definecolor{munsell}{HTML}{008FA8}
\definecolor{banana}{HTML}{FFD932}
\definecolor{cgblue}{HTML}{007CA5}
\definecolor{isabelline}{HTML}{EAEDEA}

\DeclareMathOperator{\End}{End}

\newcommand{\cat}{\mathcal{C}}
\newcommand{\ann}{\mathrm{Ann}}
\renewcommand{\Vec}{\textbf{Vec}}
\newcommand{\Z}{\mathbb{Z}}

\theoremstyle{plain}

\theoremstyle{definition}
\newtheorem{definition}{Definition}[section]

\newtheorem{example}{Example}[section]

\theoremstyle{remark}

\usepackage[foot]{amsaddr}
\makeatletter
\renewcommand{\email}[2][]{%
	\ifx\emails\@empty\relax\else{\g@addto@macro\emails{,\space}}\fi%
	\@ifnotempty{#1}{\g@addto@macro\emails{\textrm{(#1)}\space}}%
	\g@addto@macro\emails{#2}%
}
\makeatother

\title{Gauging defects in quantum spin systems}
\author{Jacob C. Bridgeman$^*$}
\address[*]{Perimeter Institute for Theoretical Physics, Waterloo, Ontario, Canada}
\email{jcbridgeman1@gmail.com}
\author{Alexander Hahn$^\dagger$}
\author{Tobias J.\ Osborne$^\dagger$}
\author{Ramona Wolf$^\dagger$}
\address[$\dagger$]{Institut für Theoretische Physik, Leibniz Universität Hannover, Hannover, Germany}
\email{alexander.hahn@htp-tel.de,\\ tobias.osborne@itp.uni-hannover.de, ramona.wolf@itp.uni-hannover.de}

\begin{document}

\begin{abstract}
The goal of this work is to build a dynamical theory of defects for quantum spin systems. A kinematic theory for an indefinite number of defects is first introduced exploiting \emph{distinguishable Fock space}. Dynamics are then incorporated by allowing the defects to become mobile via a microscopic Hamiltonian. This construction is extended to topologically ordered systems by restricting to the ground state eigenspace of Hamiltonians generalizing the \emph{golden chain}. We illustrate the construction with the example of a spin chain with $\Vec(\Z/2\Z)$ fusion rules, employing generalized tube algebra techniques to model the defects in the chain. The resulting dynamical defect model is equivalent to the critical transverse Ising model.
\end{abstract}

\maketitle

\section{Introduction}

Many important discoveries in condensed matter physics during the past decades have arisen from the study of \emph{topological states of matter} (see, for example, \cite{Wen07,NSSFD08,HK10,QZ11}). For a long time it was thought that all continuous phase transitions could be described by the Ginzburg-Landau theory of symmetry breaking. In the late 1980s, however, physicists discovered that some systems can exhibit a new kind of order going beyond the usual symmetry breaking paradigm. This new kind of order -- \emph{topological order} \cite{Wen90} -- soon became an interesting topic in its own right, useful for the description of different quantum Hall states \cite{WN90}. More recently, topologically ordered systems gave rise to \emph{topological quantum computation}, because of their remarkable properties: these locally interacting systems exhibit emergent global properties protected against environmental noise. This makes them a promising platform for the robust encoding, storage, and manipulation of quantum systems \cite{DKLP2002,Kit03,NSSFD08,Ter15,PY15,BLPSW16}.

However, in many phases, which includes those most suitable for experimental realization, the quantum computational power of a topologically ordered system is severely limited. To ameliorate this, defects in topologically ordered systems were considered \cite{RH07,Bombin2010,KK12,FSV13,BJQ13b,BASP14,JPSV15,DIP16,CCW16,BBD17,CCW17b,CCW17,BLKW17,KPEB18,ET19}, since a theory which includes defects can describe topological phases with enhanced computational power \cite{Freedman1998,FLW02b,FLW02,FKLW02}. For example, it is possible to use topologically ordered systems which are not universal for quantum computation to realize a computationally universal braiding gate set by adding defects \cite{BJQ13}.

In condensed matter physics, there is another approach to adding defects, which even in the case of simple and well-understood quantum systems substantially enriches their physics. Consider a simple example of fermions moving freely in one spatial dimension: the addition of a single defect -- an \emph{impurity} -- leads to a substantially richer and more complex system exhibiting Kondo-type effects \cite{andersonLocalizedMagneticStates1961,hewsonKondoProblemHeavy1997}. The study of such impurity models by Wilson \cite{wilsonRenormalizationGroupCritical1975} directly led to the revolutionary development of the density matrix renormalization group (DMRG) \cite{whiteDensityMatrixFormulation1992}, and the subsequent tensor network revolution \cite{bridgemanHandwavingInterpretiveDance2017}.

In the presence of additional symmetries, the phases of quantum systems have an even finer classification \cite{Wen2002,SRFL08,Kitaev2009,FK10,CGW11,FK11,TPB11,LS12,LV12,FM13,EH13,NCMT14,WPS14,K14,F14,EN14,MFCV15,BRSX15,LV16}. More precisely, it is possible for two quantum states that are equivalent when there is no additional symmetry present to be distinct in the presence of additional symmetries. This phenomenon is referred to as \emph{symmetry-protected topological} order (SPT) \cite{CGLW13,Yoshida2015,Yoshida2017} if the gapped phase is trivial in the absence of any symmetry, or as \emph{symmetry-enriched topological} (SET) order \cite{ENO10,MR13,WBV17} if the phase is topologically nontrivial. As far as defects are concerned, not only the distinct phases of matter acquire a finer classification in the presence of additional symmetries, but also the class of possible defects becomes larger. 

It is possible to to transform a topologically ordered system in the presence of a global symmetry to a system with a local gauge invariance. This is referred to as ``gauging the symmetry''. This is convenient for several reasons. For example, many properties of the ungauged system can be understood by looking at the properties of the gauged theory \cite{LG12,HW12,Swingle2014,CG14}. Furthermore, it may give insights into the quantum phase transitions between two different topological phases of matter \cite{BS09,BW10,BW11,BSS12}. 
The study of the fusion properties of defects in general quantum phases -- particularly when they are allowed to be mobile -- is deeply connected with global and local symmetries and the formalism of \emph{G-crossed braided tensor categories} \cite{BBCW14}. This connection has been studied from a mathematical point of view in various works \cite{T00,ENO10,Turaev2010,CGPW16,EMJP18,CSZW18,D19,BJ19}. The study of models involving mobile defects -- analogues of impurity models for anyons -- with nontrivial dynamics lies at the very cutting edge of current research. Such models promise to provide new phases of matter even in $(1+1)$ dimensions, going beyond the SPT classification.

In our work, the goal is to make physical and microscopic sense of a dynamical theory of defects for quantum spin systems. While it is straightforward to define the kinematical data for a quantum system with an indefinite number of defects via the well-known Fock space construction, imposing dynamics is more complicated. In the context of topologically ordered systems, dynamical information is typically introduced via the ground eigenspace of a specific Hamiltonian that represents the topological properties of the system. A common example is Kitaev's toric code for $N\times N$ quantum spins arranged on the torus, which exhibits a four-dimensional ground eigenspace. Defects in this system could be modeled by absent quantum spins, resulting in ground eigenspaces of differing dimensions. When considering more than one defect (i.e., missing spins) one also has to distinguish between cases where the defects are next to each other and those where they are spatially separated. This can result in a  complicated combinatorial problem, which we investigate in more detail in Section \ref{Gauging}.

Furthermore, we study in great detail the one-dimensional example of a $\Vec(\Z/2\Z)$ spin chain, where defects are modeled by an invertible bimodule of the category in Section \ref{subsec:VecZ2}.
Using generalized tube algebra techniques \cite{ocneanu}, we construct explicit trivalent vertices for the gauged theory. Consistent with the literature \cite{TY,ENO10,Bombin2010,BBCW14,WBV17}, we find that the $F$-symbols of the gauged theory (computed using these explicit vertices) coincide exactly with those of the Ising category. To the best of our knowledge, this is the first time such a result has been computed directly using the tube algebra. Furthermore, defining a golden-chain-like Hamiltonian \cite{Feiguin2007} for this spin chain results in the transverse Ising model.

This paper is structured as follows: Section \ref{Gauging} is dedicated to the general theoretical description of a dynamical theory of defects. We first introduce kinematics for defects in quantum spin systems before proposing a way to model dynamically evolving defects. In Section~\ref{S:defs}, we provide definitions and background required for the remainder of the paper. In particular, we need the concepts from generalized tube algebras provided here to be able to study explicit examples of models with dynamical defects. Furthermore, we present a spin chain with $\Vec(\Z/2\Z)$ fusion rules in Section \ref{Ising} and compute its dynamical data. This leads to a realization of the transverse Ising model. Finally, we conclude in Section \ref{Conclusion}.

\section{Dynamical theory of defects for quantum spin systems}
\label{Gauging}

Our goal in this section is to present a framework to make physical and microscopic sense of a dynamical theory of defects for quantum (spin) systems. We restrict our attention to defects modeled by absent/missing quantum spins. (Extending this idea to more general defects modeled by different Hilbert spaces then becomes a straightforward task.)

We build up to a theory of dynamical defects in stages. In the first stage we focus on the problem of modeling an \emph{indefinite} number of distinguishable spins. Building on this we can propose a kinematical space to model arbitrary numbers of quantum spins at various locations. This allows us to then consider the case of missing-spin defects for topologically ordered systems.

\subsection{Describing an indefinite number of distinguishable quantum spins}

When we study quantum theory we learn that to describe the Hilbert space $\mathcal{H}_{AB}$ of systems comprised of two or more distinguishable subsystems we should use the tensor product:
\begin{equation}
	\mathcal{H}_{AB} \cong \mathcal{H}_A\otimes \mathcal{H}_B,
\end{equation}
where $\mathcal{H}_A$ and $\mathcal{H}_B$ are the Hilbert spaces for subsystems $A$ and $B$, respectively. We also learn that to describe a system with an \emph{indefinite} number of \emph{indistinguishable} particles we should use \emph{Fock space}. The construction of Fock space is often presented in a confusing way which intermixes the roles of exchange symmetry and indefinite particle number.

What is not so commonly taught, but should be, is the arguably conceptually simpler construction of \emph{distinguishable Fock space} (see \cite{OsborneVideoLectureAQT} for a course where this is explained), appropriate for describing a collection of an indefinite number of distinguishable particles. Subsequently, one can obtain Bose-/Fermi-Fock space by imposing an equivalence relation under particle exchange. This approach emphasizes the separation of concerns, and, as we will see, also affords us considerable freedom in modeling more exotic situations such as the description of dynamical defects.

Suppose we want to model the following situation. Imagine we have a system comprised of $N$ distinguishable quantum spins with local Hilbert space $\mathbb{C}^d$. As per the composite-system axiom of quantum mechanics, we model such a system with the Hilbert space
\begin{equation}
	\mathcal{H}_N \cong \bigotimes_{j=1}^N \mathbb{C}^d.
\end{equation}
This Hilbert space describes all states of the $N$ \emph{individually identifiable and addressable} quantum spins. The $j$th tensor factor corresponds to the internal states of the $j$th spin.

What if we don't know beforehand how many particles we'll end up with? Let's assume that, even though we don't know how many particles we'll have, we will still be able to individually identify and address the particles\footnote{If you want a physical example for such a system: imagine a $1D$ optical lattice in which either zero or one two-level atoms can be present at each lattice site. Further, assume that the atoms line up in a contiguous line, with no gaps, i.e., we impose some additional potential gradient. Although we don't know how many particles we'll get, we can still identify and address the particles via lattice location.}. How should we incorporate the additional quantum number $N$, describing the size of the collection? The answer is via the \emph{direct sum}: we take the direct sum over $N$ of the Hilbert space $\mathcal{H}_N$ for $N$ distinguishable particles. This results in \emph{distinguishable Fock space}
\begin{equation}
	\mathfrak{F}(\mathbb{C}^d) \cong \bigoplus_{N=0}^\infty \mathcal{H}_N\cong \bigoplus_{N=0}^\infty (\underbrace{\mathbb{C}^d\otimes \mathbb{C}^d\otimes \cdots \otimes \mathbb{C}^d}_{\text{$N$ factors}}).
\end{equation}
By convention and assumption the space describing zero particles, the \emph{vacuum}, is assigned the Hilbert space $(\mathbb{C}^d)^{\otimes 0} \cong \mathbb{C}$.

It is now easy to incorporate additional constraints on the numbers of particles, e.g., to describe a system comprised of \emph{either} zero distinguishable quantum spins \emph{or} one distinguishable quantum spin we would use
\begin{equation}
	\mathfrak{F}_{\le 1} (\mathbb{C}^d) \cong \mathbb{C}\oplus \mathbb{C}^d.
\end{equation}
We could call this the Hilbert space of a ``maybe'' quantum spin.

Let's now consider the central situation for this work. How should we describe a quantum lattice of $n$ sites, where either a single quantum spin is present at a site, or not? (We call the case where a spin is absent a \emph{defect}.) According to the discussion above we should simply tensor up $N$ ``maybe'' quantum spins:
\begin{equation}
	\mathfrak{F}_{\le n}(\mathbb{C}^d) \equiv \bigotimes_{N=0}^n \mathfrak{F}_{\le 1}(\mathbb{C}^d) \cong (\mathbb{C}\oplus \mathbb{C}^d)^{\otimes N}.
\end{equation}
Expanding out the tensor factors leads to the equivalent definition
\begin{equation}
	\mathfrak{F}_{\le n}(\mathbb{C}^d) \equiv \bigoplus_{j=0}^N \mathbb{C}^{\binom{N}{j}}\otimes \mathcal{H}_j.
\end{equation}
A quick way to convince yourself of the validity of this second representation just take the dimension of both definitions. In the first case we get $\dim(\mathfrak{F}_{\le n}(\mathbb{C}^d)) = (d+1)^n$ and in the second we find
\begin{equation}
	\dim(\mathfrak{F}_{\le n}(\mathbb{C}^d)) = \sum_{j=0}^N \binom{N}{j} d^j,
\end{equation}
which is identical due to the binomial theorem. We call the space arising in the second representation
\begin{equation}
	\mathbb{C}^{\binom{N}{j}} \cong \mathcal{K}_{j};
\end{equation}
it may be interpreted as the configuration space of $j$ hard-core identical scalar particles arranged on a system of $N$ possible locations. Our total space still corresponds to \emph{distinguishable} particles; the tensor factor $\mathcal{K}_j$ takes care of identifying which locations are occupied with spins and the second tensor factor represents the actual distinguishable spins at those locations.

The Hilbert space $\mathfrak{F}_{\le n}(\mathbb{C}^d)$ supplies us with just the kinematical data to describe a system of distinguishable particles. To incorporate additional dynamical information we must specify additional \emph{dynamical} data.

\subsection{Modeling dynamically evolving defects}
We specify dynamics in quantum mechanics most directly via a one-parameter family of unitary operators $U_t:\mathcal{H}\rightarrow\mathcal{H}$; usually the family is continuous enough that we can represent them via a \emph{Hamiltonian} $H$:
\begin{equation}
	U_t = e^{-it H}.
\end{equation}
To introduce dynamics for our defect system we hence just need a hermitian matrix acting on $\mathfrak{F}_{\le n}(\mathbb{C}^d)$. This is easy enough in the abstract, however, particularly for topologically ordered systems we introduce dynamical information more indirectly.

In the context of topologically ordered models such as Kitaev's toric code \cite{Kit03} we typically introduce dynamical information indirectly by describing the system via the ground eigenspace $\mathcal{V}\subset \mathcal{H}$ of a specific Hamiltonian $H$. In the case of the toric code for $N\times N$ quantum spins arranged on the torus, the ground eigenspace $\mathcal{V}_{\mathbb{T}}$ of the toric code Hamiltonian is four dimensional.

\begin{figure}
	\centering
	\begin{subfigure}[t]{0.2\linewidth}
		\centering
		\begin{tikzpicture}[scale=0.75,baseline=(current bounding box.center)]
			\draw (-0.25,0) -- (3.25,0);
			\draw (-0.25,-1) -- (3.25,-1);
			\draw (-0.25,-2) -- (1,-2);
			\draw (2,-2) -- (3.25,-2);
			\draw (-0.25,-3) -- (3.25,-3);
			\draw (0,-3.25) -- (0,0.25);
			\draw (1,-3.25) -- (1,0.25);
			\draw (2,-3.25) -- (2,0.25);
			\draw (3,-3.25) -- (3,0.25);
		\end{tikzpicture}
		\caption{One missing qubit defect on a toric code lattice.}\label{fig:1defect}
	\end{subfigure}\hspace{40pt}
	\begin{subfigure}[t]{0.2\linewidth}
		\centering
		\begin{tikzpicture}[scale=0.75,baseline=(current bounding box.center)]
			\draw (-0.25,0) -- (3.25,0);
			\draw (-0.25,-1) -- (1,-1);
			\draw (2,-1) -- (3.25,-1);
			\draw (-0.25,-2) -- (1,-2);
			\draw (2,-2) -- (3.25,-2);
			\draw (-0.25,-3) -- (3.25,-3);
			\draw (0,-3.25) -- (0,0.25);
			\draw (1,-3.25) -- (1,0.25);
			\draw (2,-3.25) -- (2,0.25);
			\draw (3,-3.25) -- (3,0.25);
		\end{tikzpicture}
		\caption{Two adjacent missing qubit defects on a toric code lattice.}\label{fig:2adjdefects}
	\end{subfigure}\hspace{40pt}
	\begin{subfigure}[t]{0.2\linewidth}
		\centering
		\begin{tikzpicture}[scale=0.75,baseline=(current bounding box.center)]
		\draw (-0.25,0) -- (3.25,0);
		\draw (-0.25,-1) -- (2,-1);
		\draw (3,-1) -- (3.25,-1);
		\draw (-0.25,-2) -- (1,-2);
		\draw (2,-2) -- (3.25,-2);
		\draw (-0.25,-3) -- (3.25,-3);
		\draw (0,-3.25) -- (0,0.25);
		\draw (1,-3.25) -- (1,0.25);
		\draw (2,-3.25) -- (2,0.25);
		\draw (3,-3.25) -- (3,0.25);
		\end{tikzpicture}
		\caption{Two distant missing qubit defects on a toric code lattice.}\label{fig:2distdefects}
	\end{subfigure}
	\caption{Depiction of different situations where missing qubit defects are inserted on a toric code lattice. In the case of more than one missing link the ground space for the system depends on whether the defects are adjacent to each other.}
\end{figure}
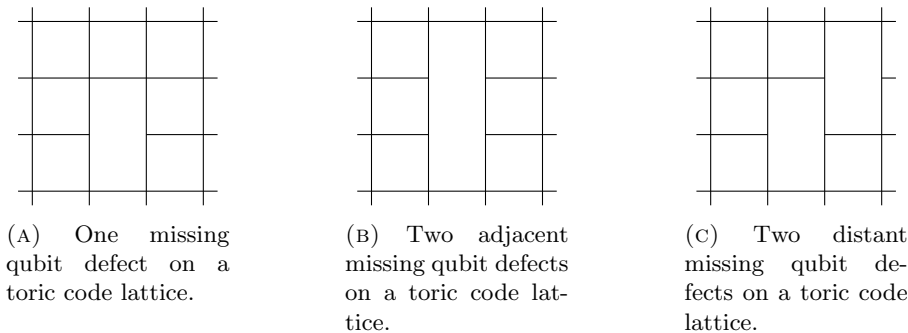

Continuing in the context of the toric code, let's suppose we now have a lattice with a missing-qubit defect at some edge $e$ \cite{BLKW17}, see Figure \ref{fig:1defect}. One can define a toric-code Hamiltonian for this new punctured torus; restricting to its ground eigenspace yields a \emph{four-dimensional subspace} $\mathcal{V}_{\mathbb{T}\setminus e}$. There is no obstruction to describing the ground eigenspace for two, three, etc.\ missing-qubit defects. What results is a rather intricate combinatorial problem, as, depending on where the defects are located relative to each other, one gets higher or lower dimensional ground eigenspace. To see an example of the intricacies that easily result, consider the case of two missing edges: these can either be adjacent to each other or distant. In the latter case, depicted in Figure \ref{fig:2distdefects}), the ground space for this system is $(\mathbb{C}^2)^{\otimes 3}$ while in the case of adjacent defects we have a single larger puncture (with smooth boundary), see Figure \ref{fig:2adjdefects}, and the ground eigenspace is only $(\mathbb{C}^2)^{\otimes 2}$.

We can incorporate such \emph{indirect} dynamical information into our kinematical space $\mathcal{F}_{\le n}(\mathbb{C}^d)$ as follows. We start by writing out 
\begin{equation}
	\mathcal{V}_{\le n}(\mathbb{C}^d) = \bigoplus_{e_1,e_2, \ldots, e_{N^2} =0}^1 \mathcal{V}_{e_1,e_2,\ldots, e_N^2},
\end{equation}
where $\mathcal{V}_{e_1,e_2,\ldots, e_N^2}$ is the ground eigenspace for the toric code with a missing-edge defect at every edge $e_j$ with $e_j=0$. Writing out the first terms of this big direct sum gives us
\begin{equation}
	\mathcal{V}_{\le n}(\mathbb{C}^d) = (\mathbb{C}^2\otimes \mathbb{C}^2)\oplus \mathbb{C}^{N^2}\otimes (\mathbb{C}^2\otimes \mathbb{C}^2\otimes \mathbb{C}^2)\oplus \cdots.
\end{equation}
The first direct summand corresponds to the ground eigenspace of the toric code without defects, the second summand corresponds to the $N^2$ possible single defects, etc.

The full direct-sum structure of $\mathcal{V}_{\le n}(\mathbb{C}^d)$ is not easy to describe. Indeed, this is why we use a category theoretic language to discuss these problems. 

So far, the space $\mathcal{V}_{\le n}(\mathbb{C}^d)$ is the low-energy configuration space for a topologically ordered system, the toric code, on a lattice with an indefinite number of missing-edge defects. In this case, we assume we somehow \emph{can know} where the defects are located. However, for a full dynamical theory, this assumption is potentially unjustified.

To incorporate the effects of such limited detection ability we impose an \emph{equivalence relation} on $\mathcal{V}_{\le n}(\mathbb{C}^d)$ where we identify ``physically indistinguishable'' configurations of defects. There are many possible notions of indistinguishability: this is an \emph{operational} notion and must be justified on a case by case basis. 

One particular, rather coarse, notion of indistinguishability may be justified as follows. Suppose we have a state $|\phi\rangle$ of the system for a defect configuration $e_1,e_2, \ldots, e_{N^2}$ and another $|\psi\rangle$ for a defect configuration with the same number of defects at possible different locations $e_1',e_2',\ldots, e'_{N^2}$. We say that the two states $|\phi\rangle$ and $|\psi\rangle$ are \emph{equivalent} if there is a unitary circuit which can transform $|\phi\rangle$ to $|\psi\rangle$. This, in particular, requires that the subspaces $\mathcal{V}_{e_1,e_2,\ldots,e_{N^2}}$ and $\mathcal{V}_{e_1',e_2',\ldots,e_{N^2}'}$ which contain $|\phi\rangle$ and $|\psi\rangle$ have the same dimensions.

This equivalence relation collapses many of the direct sums appearing in $\mathcal{V}_{e_1,e_2,\ldots,e_{N^2}}$, e.g., after applying the equivalence relation the first two summands above become
\begin{equation}
	\left(\mathcal{V}_{\le n}(\mathbb{C}^d)/\sim\right) = (\mathbb{C}^2\otimes \mathbb{C}^2)\oplus (\mathbb{C}^2\otimes \mathbb{C}^2\otimes \mathbb{C}^2)\oplus \cdots.
\end{equation}

Writing out the whole ground eigenspace and applying the equivalence relation to it results in a highly intricately combinatorial problem, which is why we exploit concepts from category theory to model defects in the system. Of course, the above description is not only restricted to the toric code case. In the next section, we show how this construction works for a one-dimensional spin system and in Section~\ref{Ising}, we show ideas and techniques from category theory can be used to model dynamical defects in such a one-dimensional spin system.


\subsection{Example: Spin chain}
\label{subsec:VecZ2}

We consider a one-dimensional spin chain of $N$ particles, where each particle can be in the spin-up state or in the spin-down state, i.e., a chain
	\begin{figure}[H]
		\begin{tikzpicture}[scale=1.5]
			\fill[black] (0,0) circle (0.07cm);
			\fill[black] (0.5,0) circle (0.07cm);
			\fill[black] (1,0) circle (0.07cm);
			\fill[black] (1.5,0) circle (0.07cm);
			\fill[black] (2,0) circle (0.07cm);
			\fill[black] (2.5,0) circle (0.07cm);
			\fill[black] (3,0) circle (0.07cm);
			\fill[black] (4,0) circle (0.07cm);
			\node at (3.5,0) {$\dots$};
		\end{tikzpicture}
	\end{figure}
\noindent
whose Hilbert space is 
	\begin{equation}
		\mathcal{H}_0=\bigotimes_{i=1}^N \mathbb{C}^2.
	\end{equation}
We now introduce defects, which means that one of the spins, say the spin at site $j$, is replaced by a different kind of spin, e.g., no spin (indicated in red): 
	\begin{figure}[H]
		\begin{tikzpicture}[scale=1.5]
			\fill[black] (0,0) circle (0.07cm);
			\fill[black] (0.5,0) circle (0.07cm);
			\fill[black] (1,0) circle (0.07cm);
			\fill[black] (1.5,0) circle (0.07cm);
			\fill[red] (2,0) circle (0.07cm);
			\fill[black] (2.5,0) circle (0.07cm);
			\fill[black] (3,0) circle (0.07cm);
			\fill[black] (4,0) circle (0.07cm);
			\node at (3.5,0) {$\dots$};
		\end{tikzpicture}
	\end{figure}
\noindent
This corresponds to replacing the Hilbert space $\mathbb{C}^2$ at site $j$ with $\mathbb{C}$, which results in the overall Hilbert space
	\begin{equation}
		\mathcal{H}_1^{(j)}=\left(\bigotimes_{i=1}^{j-1}\mathbb{C}^2\right)\otimes\mathbb{C}\otimes\left(\bigotimes_{i=j+1}^N\mathbb{C}^2\right).
	\end{equation}
The subscript here denotes the number of defects in the chain and the superscript indicates at which site the defect appears. If we want to consider both possibilities, having no defect and having a defect at site $j$, we use the Hilbert space
	\begin{equation}
		\mathcal{H}=\mathcal{H}_0\oplus\mathcal{H}_1^{(j)}.
	\end{equation}
We can also allow the defect to move, which means we still restrict the setting to only one defect in total, but it can happen at any site. Hence, the overall Hilbert space becomes
	\begin{equation}
		\mathcal{H}=\mathcal{H}_0\oplus\left(\bigoplus_{j=1}^N\mathcal{H}_1^{(j)}\right).
	\end{equation}
This construction can be generalized to an arbitrary number of defects: The Hilbert space for having defects two defects in the chain, say at sites $j$ and $k$, is 
	\begin{equation}
		\mathcal{H}_2^{(j,k)}=\left(\bigotimes_{i=1}^{j-1}\mathbb{C}^2\right)\otimes\mathbb{C}\otimes\left(\bigotimes_{i=j+1}^{k-1}\mathbb{C}^2\right)\otimes\mathbb{C}\otimes\left(\bigotimes_{i=k+1}^{N}\mathbb{C}^2\right).
	\end{equation}
Again, if we allow these defects to move and also include the possibilities of having only one defect and no defect at all, the overall Hilbert space is
	\begin{equation}
		\mathcal{H}=\mathcal{H}_0\oplus\left(\bigoplus_{j=1}^N\mathcal{H}_1^{(j)}\right)\oplus\left(\bigoplus_{j=1}^N\bigoplus_{k\neq j}\mathcal{H}_2^{(j,k)}\right).
	\end{equation}
We can continue this construction until we have a defect at every site of the chain, i.e.
	\begin{equation}
		\mathcal{H}_N=\bigotimes_{i=1}^N\mathbb{C},
	\end{equation}
and the overall Hilbert space is then
	\begin{equation}
		\mathcal{H}=\bigoplus_{n\in\#\mathrm{defects}}\mathcal{H}_n,
	\end{equation}
where $\mathcal{H}_n$ is the direct sum of all possible Hilbert spaces with $n$ defects, as constructed above. Since this is now a very complicated Hilbert space, we can also think about it in a different and simpler way: at each site, the particle can be in one of three states: spin up, spin down, or no spin. Hence, we have effectively a three-level system at each site of the chain, and therefore the overall Hilbert space can be written as
	\begin{equation}
		\mathcal{H}\cong\bigotimes_{j=1}^N\left(\mathbb{C}\oplus\mathbb{C}^2\right).
	\end{equation}

We study a concrete example in detail in Section~\ref{Ising}, namely a particle chain in the sense of the golden chain presented in \cite{Feiguin2007}, which is based on a fusion category with $\Vec(\Z/2\Z)$ fusion rules with pairwise interaction. Since the discussion of this example requires some background on fusion categories and bimodules as well as annular categories, we present the mathematical preliminaries in the next section before explicitly constructing the defects afterwards.

\section{Background}\label{S:defs}
In this section, we provide all the definitions and notation in the vein of category theory that are used throughout the subsequent computations. We do not give exhaustive details and explanations whenever it is not necessary for this work but refer to the literature where those can be found.


\subsection{Fusion categories}

We briefly recall the definition of a fusion category, omitting many details. A full definition can be found in Chapter 4 of \cite{Etingof2015}.
\begin{definition}[Fusion category]
	A tensor category $\cat$ is a $\mathbb{C}$-linear category $\cat$ with a functor $\otimes:\cat\times\cat\to\cat$, a natural isomorphism $(-\otimes-)\otimes-\cong-\otimes(-\otimes-)$ called the associator, and a unit object $1$.
	These data must satisfy coherence equations such as the pentagon equations, that can be found in Section 2.1 of \cite{Etingof2015}.
	
	A \emph{fusion category} is a rigid\footnote{We refer the reader to \cite{Etingof2015} for details.} semisimple tensor category with finitely many isomorphism classes of simple objects and $\End(1)\cong \mathbb{C}$.
	
	It is convenient to describe a fusion category $\cat$ using \emph{skeletal data}. This approach is standard in the (physics) literature, and amounts to picking representatives for each isomorphism class of simple object. The full fusion category can be canonically reconstructed from the skeletal data \cite{BBJSkeletal}. A \emph{skeletal fusion category} consists of:
	\begin{itemize}
		\item A finite set of simple objects $L_\mathcal{C}=\{1,a,b,c,\ldots\}$ , where $1$ is the distinguished unit object.
		\item For each triple $(a,b;c)$ a finite dimensional Hilbert space $C(a\otimes b,c)$ represented as a trivalent vertex
		\begin{equation}
		\begin{tikzpicture}[scale=0.8,baseline=(current bounding box.center)]
		\draw (-.707,-.707)--(0,0) node[pos=-.25] {$a$};
		\draw (.707,-.707)--(0,0) node[pos=-.25] {$b$};
		\draw (0,0)--(0,1) node[pos=1.25] {$c$};
		\end{tikzpicture}.
		\end{equation} 
		\item An associator. If a basis is picked for each $C(-\otimes -,-)$ (here we assume these are all 1 dimensional for simplicity), the associator can be represented as a tensor $F$
		\begin{equation}
		\begin{tikzpicture}[scale=0.8,baseline=(current bounding box.center)]
		\draw (0,0)--(0,1) node[pos=1.25] {$d$};
		\draw (0,0)--(1.41,-1.41) node[pos=1.15] {$c$};
		\draw (0,0)--(-.707,-.707) node [pos=.5,above left] {$e$};
		\draw (-.707,-.707)--(-1.41,-1.41)node[pos=1.25] {$a$};
		\draw (-.707,-.707)--(0,-1.41)node[pos=1.25] {$b$};
		\end{tikzpicture}
		=\sum_{f\in L}\left(F_{abc}^{d}\right)_{e,f}
		\begin{tikzpicture}[scale=0.8,baseline=(current bounding box.center)]
		\draw (0,0)--(0,1) node[pos=1.25] {$d$};
		\draw (.707,-.707)--(1.41,-1.41) node[pos=1.25] {$c$};
		\draw (0,0)--(.707,-.707) node [pos=.5,above right] {$f$};
		\draw (0,0)--(-1.41,-1.41)node[pos=1.15] {$a$};
		\draw (.707,-.707)--(0,-1.41)node[pos=1.25] {$b$};
		\end{tikzpicture}.\label{eqn:Fsymbdef}
		\end{equation} 
		The tensors are often called the $F$-symbols.
	\end{itemize}
\end{definition}

\begin{example}[$\Vec(G)$]\label{example:vecG}
	Let $G$ be a finite group. The category $\Vec(G)$ is the category of $G$-graded vector spaces. The skeletal description is very simple. The set of simples is $G$ as a set, with the group identity becoming the unit. The vector spaces are 0 dimensional except $C(g\otimes h, gh)$ which is 1 dimensional for all $g,h$. The $F$-symbols are $+1$ when the diagrams in \eqref{eqn:Fsymbdef} are nonzero.
\end{example}

\subsection{Bimodules}

\begin{definition}
	Let $\mathcal{C}$ be a fusion category. A \textbf{left module category} over $\mathcal{C}$ is a semi-simple category $\mathcal{M}$ equipped with a left $\mathcal{C}$-\emph{action}, i.e.\ a functor $\triangleright:\mathcal{C}\times\mathcal{M}\to\mathcal{M}$ and a natural isomorphism 
		\begin{equation}
			a\triangleright(b\triangleright m)\cong(a\otimes b)\triangleright m
		\end{equation}
	which satisfies some coherence conditions that can be found in Section~7.1 of \cite{Etingof2015}. If we choose bases for all the vector spaces $\mathcal{M}(a\triangleright m,n)$, this associator can be written in terms of a tensor $L$ using string diagrams
		\begin{equation}
			\begin{tikzpicture}[scale=0.9,baseline=(current bounding box.center)]
			\node[red](m) at (0,0) {$m$};
			\node(b) at (-1,0) {$b$};
			\node(a) at (-2,0) {$a$};
			\node[red](abm) at (0,3) {$n$};
			\draw[red,line width=0.4mm] (m) to node[right] {$p$} (abm);
			\draw (b) to [bend left] (0,1);
			\draw (a) to [bend left] (0,2);
			\end{tikzpicture}\ =\sum_q \left(L^n_{abm}\right)_{p,q}
			\begin{tikzpicture}[scale=0.9,baseline=(current bounding box.center)]
			\node[red](m) at (0,0) {$m$};
			\node(b) at (-1,0) {$b$};
			\node(a) at (-2,0) {$a$};
			\node[red](abm) at (0,3) {$n$};
			\node(ab) at (-1.2,1.8) {$q$};
			\draw[red,line width=0.4mm] (m) to (abm);
			\draw (b) to [bend right] (-1.5,1.1);
			\draw (a) to [bend left] (0,2);
			\end{tikzpicture},\label{eqn:L}
		\end{equation}
	where we have marked the objects of the module category in red. One can define a \textbf{right module category} over $\mathcal{C}$ in a similar way: Here, we have a semi-simple category $\mathcal{M}$ with a functor $\triangleleft:\mathcal{M}\times\mathcal{C}\to\mathcal{M}$ and a natural isomorphism
		\begin{equation}
			(m\triangleleft a)\triangleleft b\cong m\triangleleft(a\otimes b)
		\end{equation}
	(satisfying some coherence conditions), which we can represent by a tensor $R$:
		\begin{equation}
			\begin{tikzpicture}[scale=0.9,baseline=(current bounding box.center)]
			\node[red](m) at (0,0) {$m$};
			\node(b) at (1,0) {$a$};
			\node(a) at (2,0) {$b$};
			\node[red](abm) at (0,3) {$n$};
			\draw[red,line width=0.4mm] (m) to node [left] {$p$} (abm);
			\draw (b) to [bend right] (0,1);
			\draw (a) to [bend right] (0,2);
			\end{tikzpicture}\ =\sum_q\left(R^n_{mab}\right)_{p,q}
			\begin{tikzpicture}[scale=0.9,baseline=(current bounding box.center)]
			\node[red](m) at (0,0) {$m$};
			\node(b) at (1,0) {$a$};
			\node(a) at (2,0) {$b$};
			\node[red](abm) at (0,3) {$n$};
			\node(ab) at (1.2,1.8) {$q$};
			\draw[red,line width=0.4mm] (m) to (abm);
			\draw (b) to [bend left] (1.5,1.1);
			\draw (a) to [bend right] (0,2);
			\end{tikzpicture}.\label{eqn:R}
		\end{equation}
	
\begin{definition}
	Let $\mathcal{C}, \mathcal{D}$ be fusion categories. A $(\mathcal{C},\mathcal{D})$-\textbf{bimodule category} (also written $\mathcal{C}\curvearrowright\mathcal{M}\curvearrowleft\mathcal{D}$) is a semi-simple category $\mathcal{M}$ that has left $\mathcal{C}$-module and right $\mathcal{D}$-module category structures, i.e.\ there are natural isomorphisms 
		\begin{align}
			a\triangleright(b\triangleright m)&\cong(a\otimes b)\triangleright m\\ 
			(m\triangleleft a)\triangleleft b&\cong m\triangleleft(a\otimes b)
		\end{align}
	that satisfy certain coherence conditions between $L$ and $R$. Additionally, there is a natural isomorphism 
		\begin{equation}
			a\triangleright(m\triangleleft b)\cong (a\triangleright m)\triangleleft b.
		\end{equation}
	As above, after choosing bases for the morphism spaces, this can be represented in terms of string diagrams with a tensor $C$:
		\begin{equation}
			\begin{tikzpicture}[scale=0.9,baseline=(current bounding box.center)]
			\node[red](m) at (0,0) {$m$};
			\node(a) at (-1,0) {$a$};
			\node(b) at (1,0) {$b$};
			\node[red](abm) at (0,3) {$n$};
			\draw[red,line width=0.4mm] (m) to node [left] {$p$} (abm);
			\draw (a) to [bend left] (0,1);
			\draw (b) to [bend right] (0,2);
			\end{tikzpicture}\ =\sum_q\left(C^n_{amb}\right)_{pq}
			\begin{tikzpicture}[scale=0.9,baseline=(current bounding box.center)]
			\node[red](m) at (0,0) {$m$};
			\node(a) at (-1,0) {$a$};
			\node(b) at (1,0) {$b$};
			\node[red](abm) at (0,3) {$n$};
			\draw[red,line width=0.4mm] (m) to node [right] {$q$} (abm);
			\draw (b) to [bend right] (0,1);
			\draw (a) to [bend left] (0,2);
			\end{tikzpicture}.
		\end{equation}
\end{definition}

\end{definition}

\subsection{Example: $\Vec(\Z/p\Z)$ bimodules}\label{sec:VecZp-bimodules}

We now list all the $\Vec(\Z/p\Z)$-$\Vec(\Z/p\Z)$ bimodules since we are going to need them in Subsection \ref{subsec:VecZ2}. This data is mostly taken from \cite{BBJ19}, and the names assigned to the bimodules are taken from there. Each bimodule $\mathcal{M}$ is labeled by a conjugacy class of subgroups $H\subseteq\Z/p\Z\times\Z/p\Z$ and a $2-$cocycle $\omega\in H^2(H,U(1))$ (see \cite{Etingof2015}). The objects of $\mathcal{M}$ are labeled by cosets of $H$. In general, there are four non-invertible bimodules: 
	\begin{enumerate}
		\item The trivial bimodule $T$, which is coming from the subgroup $\{(0,0)\}$. It has $p^2$ simple objects, which are labeled by the cosets $\{(g,h)\}$ of the group, i.e.\ a simple object in $T$ is $(g,h)$. The left and right action is given as follows
			\begin{equation}
				\begin{tikzpicture}[scale=0.8,baseline=(current bounding box.center)]
				\node[red](m) at (0,0) {$(g,h)$};
				\node(b) at (-1,0) {$a$};
				\node[red](abm) at (0,2) {$(g+a,h)$};
				\draw[red,line width=0.4mm] (m)-- (abm);
				\draw (b) to [bend left] (0,1);
				\end{tikzpicture}\hspace{20pt}
				\begin{tikzpicture}[scale=0.8,baseline=(current bounding box.center)]
				\node[red](m) at (0,0) {$(g,h)$};
				\node(b) at (1,0) {$a$};
				\node[red](abm) at (0,2) {$(g,h+a)$};
				\draw[red,line width=0.4mm] (m)-- (abm);
				\draw (b) to [bend right] (0,1);
				\end{tikzpicture}
			\end{equation}
		\noindent
		and the associator for this bimodule is trivial:
			\begin{equation}
				\begin{tikzpicture}[scale=0.8,baseline=(current bounding box.center)]
				\node[red](m) at (0,0) {$(g,h)$};
				\node(a) at (-1,0) {$a$};
				\node(b) at (1,0) {$b$};
				\node[red](abm) at (0,3) {$(g+a,h+b)$};
				\draw[red,line width=0.4mm] (m) -- (abm);
				\draw (a) to [bend left] (0,1);
				\draw (b) to [bend right] (0,2);
				\end{tikzpicture}\ =
				\begin{tikzpicture}[scale=0.9,baseline=(current bounding box.center)]
				\node[red](m) at (0,0) {$(g,h)$};
				\node(a) at (-1,0) {$a$};
				\node(b) at (1,0) {$b$};
				\node[red](abm) at (0,3) {$(g+a,h+b)$};
				\draw[red,line width=0.4mm] (m) -- (abm);
				\draw (b) to [bend right] (0,1);
				\draw (a) to [bend left] (0,2);
				\end{tikzpicture}.
			\end{equation}
		\item The bimodule $L$ from the subgroup $H=\langle(1,0)\rangle\cong \Z/p\Z$, which has $p$ simple objects. The labels are given by the cosets $\{(h,g)|h\in\Z/p\Z\}$, hence the label for a simple object in $L$ is $g$. Left and right action are given by 
			\begin{equation}
				\begin{tikzpicture}[scale=0.8,baseline=(current bounding box.center)]
				\node[red](m) at (0,0) {$g$};
				\node(b) at (-1,0) {$a$};
				\node[red](abm) at (0,2) {$g$};
				\draw[red,line width=0.4mm] (m)-- (abm);
				\draw (b) to [bend left] (0,1);
				\end{tikzpicture}\hspace{20pt}
				\begin{tikzpicture}[scale=0.8,baseline=(current bounding box.center)]
				\node[red](m) at (0,0) {$g$};
				\node(b) at (1,0) {$a$};
				\node[red](abm) at (0,2) {$g+a$};
				\draw[red,line width=0.4mm] (m)-- (abm);
				\draw (b) to [bend right] (0,1);
				\end{tikzpicture}
			\end{equation}
		and the associator is trivial.
		\item The bimodule $R$ from the subgroup $H=\langle(0,1)\rangle\cong \Z/p\Z$. This bimodule is basically the same as $L$, but in all operations left and right are flipped.
		\item $F_0$, the bimodule from the subgroup $\langle(0,1),(1,0)\rangle\cong\Z/p\Z\times\Z/p\Z$, which has only one object that is denoted $*$. Left and right actions are
			\begin{equation}
				\begin{tikzpicture}[scale=0.8,baseline=(current bounding box.center)]
				\node[red](m) at (0,0) {$*$};
				\node(b) at (-1,0) {$a$};
				\node[red](abm) at (0,2) {$*$};
				\draw[red,line width=0.4mm] (m)-- (abm);
				\draw (b) to [bend left] (0,1);
				\end{tikzpicture}\hspace{20pt}
				\begin{tikzpicture}[scale=0.8,baseline=(current bounding box.center)]
				\node[red](m) at (0,0) {$*$};
				\node(b) at (1,0) {$a$};
				\node[red](abm) at (0,2) {$*$};
				\draw[red,line width=0.4mm] (m)-- (abm);
				\draw (b) to [bend right] (0,1);
				\end{tikzpicture}
			\end{equation}
		\noindent
		and the associator is trivial.
	\end{enumerate}
Furthermore, there are several invertible bimodules:
	\begin{enumerate}
		\item The bimodule $X_k$, coming from the subgroup $\{(-k,1)\}\cong\Z/p\Z$ with $p$ simple objects. The cosets of this group are $\{n(-k,1)+(h,0)|n\in\Z/p\Z\}$, hence the object labels are $h$. Left and right action is given by 
			\begin{equation}
				\begin{tikzpicture}[scale=0.8,baseline=(current bounding box.center)]
				\node[red](m) at (0,0) {$h$};
				\node(b) at (-1,0) {$a$};
				\node[red](abm) at (0,2) {$h+a$};
				\draw[red,line width=0.4mm] (m)-- (abm);
				\draw (b) to [bend left] (0,1);
				\end{tikzpicture}\hspace{20pt}
				\begin{tikzpicture}[scale=0.8,baseline=(current bounding box.center)]
				\node[red](m) at (0,0) {$h$};
				\node(b) at (1,0) {$a$};
				\node[red](abm) at (0,2) {$h+ka$};
				\draw[red,line width=0.4mm] (m)-- (abm);
				\draw (b) to [bend right] (0,1);
				\end{tikzpicture}
			\end{equation}
		\noindent
		and the associator is again trivial.
		\item The bimodule $F_q$, with $q\in H^2(\Z/p\Z,U(1))\cong\Z/q\Z$. As $F_0$, it comes from the subgroup $\langle(0,1),(1,0)\rangle$ and has only one object denoted $*$. Also, left and right action are defined in the same way as they are in $F_0$, but now the associator is non-trivial:
			\begin{equation}
				\begin{tikzpicture}[scale=0.8,baseline=(current bounding box.center)]
				\node[red](m) at (0,0) {$*$};
				\node(a) at (-1,0) {$a$};
				\node(b) at (1,0) {$b$};
				\node[red](abm) at (0,3) {$*$};
				\draw[red,line width=0.4mm] (m) -- (abm);
				\draw (a) to [bend left] (0,1);
				\draw (b) to [bend right] (0,2);
				\end{tikzpicture}\ =e^{\frac{2\pi i}{p}qab}
				\begin{tikzpicture}[scale=0.9,baseline=(current bounding box.center)]
				\node[red](m) at (0,0) {$*$};
				\node(a) at (-1,0) {$a$};
				\node(b) at (1,0) {$b$};
				\node[red](abm) at (0,3) {$*$};
				\draw[red,line width=0.4mm] (m) -- (abm);
				\draw (b) to [bend right] (0,1);
				\draw (a) to [bend left] (0,2);
				\end{tikzpicture}.
			\end{equation}
		We are especially interested in the $\Vec(\Z/2\Z)$-$\Vec(\Z/2\Z)$ bimodule $F_1$ which we use to introduce defects to a $\Vec(\Z/2\Z)$ spin chain in Section \ref{Ising}. In this case the associator is given by
			\begin{equation}
				\begin{tikzpicture}[scale=0.8,baseline=(current bounding box.center)]
				\node[red](m) at (0,0) {$*$};
				\node(a) at (-1,0) {$a$};
				\node(b) at (1,0) {$b$};
				\node[red](abm) at (0,3) {$*$};
				\draw[red,line width=0.4mm] (m) -- (abm);
				\draw (a) to [bend left] (0,1);
				\draw (b) to [bend right] (0,2);
				\end{tikzpicture}\ =(-1)^{ab}
				\begin{tikzpicture}[scale=0.9,baseline=(current bounding box.center)]
				\node[red](m) at (0,0) {$*$};
				\node(a) at (-1,0) {$a$};
				\node(b) at (1,0) {$b$};
				\node[red](abm) at (0,3) {$*$};
				\draw[red,line width=0.4mm] (m) -- (abm);
				\draw (b) to [bend right] (0,1);
				\draw (a) to [bend left] (0,2);
				\end{tikzpicture}.\label{eqn:F1}
			\end{equation}
	\end{enumerate}


\subsection{Bimodule tensor products}\label{sec:bimodtensor}

To extend a category $\cat$ to include bimodule objects, we need to introduce `bimodule trivalent vertices'. To construct these, we need a product on bimodules. 

\begin{definition}[Relative tensor product]
	Given bimodules $\mathcal{A}\curvearrowright\mathcal{M}\curvearrowleft\mathcal{B}\curvearrowright{\mathcal{N}}\curvearrowleft\mathcal{C}$, the \emph{relative tensor product} $\mathcal{M}\otimes_\mathcal{B}\mathcal{N}$ has objects $(m,n)\in \mathcal{M}\otimes\mathcal{N}$, along with isomorphisms $\beta:(m\triangleleft b,n)\cong(m,b\triangleright n)$. The isomorphisms should be compatible with the module structure, for example
	\begin{align}
	\begin{tikzpicture}[scale=.98,baseline=(current bounding box.center)]
	\node (A) at (-1,0) {$((m\triangleleft b_1)\triangleleft b_2,n)$};
	\node (B) at (2,1) {$(m\triangleleft b_1,b_2\triangleright n)$};
	\node (C) at (6,1) {$(m,b_1\triangleright(b_2\triangleright n))$};
	\node (D) at (10,0) {$(m,(b_1\otimes b_2)\triangleright n)$};
	\node (E) at (4,-1) {$m\triangleleft (b_1\otimes b_2), n)$};
	\draw [-stealth,above left] (A)--(B) node [pos=.5] {$\beta$};
	\draw [-stealth,above] (B)--(C) node [pos=.5] {$\beta$};
	\draw [-stealth,above right] (C)--(D) node [pos=.5] {$L^\mathcal{N}$};
	\draw [-stealth,below left] (A)--(E) node [pos=.5] {$R^{\mathcal{M}}$};
	\draw [-stealth,below right] (E)--(D) node [pos=.5] {$\beta$};
	\end{tikzpicture},
	\end{align}
	commutes. Here, $L^\mathcal{N}$ denotes the associator for the left action in $\mathcal{N}$ and $R^\mathcal{M}$ denotes the associator for the right action in $\mathcal{M}$. Morphisms in $\mathcal{M}\otimes_\mathcal{B}\mathcal{N}$ are morphisms in $\mathcal{M}\otimes\mathcal{N}$ that are compatible with $\beta$. $\mathcal{M}\otimes_\mathcal{B}\mathcal{N}$ is an $\mathcal{A}$-$\mathcal{C}$ bimodule, and can be decomposed into simple bimodules $\mathcal{P}$  $$\mathcal{M}\otimes_\mathcal{B}\mathcal{N}\cong\oplus_\mathcal{P}N_{\mathcal{M},\mathcal{N}}^\mathcal{P}\mathcal{P},$$
	where the $N_{\mathcal{M,N}}^\mathcal{P}$ are the corresponding coefficients in the decomposition. A complete definition can be found in \cite{DSPS14}.
\end{definition}

If a given bimodule $\mathcal{P}$ occurs in the decomposition of $\mathcal{M}\otimes_\mathcal{B}\mathcal{N}$, it is natural to introduce bimodule trivalent vertices of the form
\begin{align}
\begin{tikzpicture}[baseline=(current bounding box.center)]
\node (A) at (-.707,-.707) {$m\in\mathcal{M}$};
\node (B) at (.707,-.707) {$n\in\mathcal{N}$};
\node (C) at (0,1) {$p\in\mathcal{P}$};
\draw[violet,line width=0.4mm](A)--(0,0);
\draw[brown,line width=0.4mm](B)--(0,0);
\draw[nicegreen,line width=0.4mm](0,0)--(C);
\end{tikzpicture}.
\end{align}
This is the essence of the `inflation trick' developed in \cite{BBJ18,BB19a,BB19b}. We refer the interested reader to these references for more details. 

\subsection{Definition of the annular category}

%
The first step in our gauging process for a collection of bimodules $\{\mathcal{M}_i\}$ is to extend the fusion category $\cat$ to include the objects in $\mathcal{M}_i$. From Section~\ref{sec:bimodtensor}, we have fusion rules for bimodules $\mathcal{M}\otimes\mathcal{N}=\oplus\mathcal{P}$. We utilize the \emph{annular category} to obtain trivalent vertices for these processes. 

\begin{definition}[3-string annular category]
	Let $\mathcal{A},\mathcal{B}$ and $\mathcal{C}$ be fusion categories, and $\mathcal{A}\curvearrowright\mathcal{M}\curvearrowleft\mathcal{B}\curvearrowright\mathcal{N}\curvearrowleft\mathcal{C}$ and $\mathcal{A}\curvearrowright\mathcal{P}\curvearrowleft\mathcal{C}$ bimodule categories. The 3-string annular category $\ann_{\mathcal{M},\mathcal{N};\mathcal{P}}(\mathcal{A},\mathcal{B},\mathcal{C})$ is defined as follows:
	The simple objects are triples $(m,n;p)\in\mathcal{M}\times\mathcal{N}\times\mathcal{P}$. A basis for the morphism space $(m,n;p)\to (m^\prime,n^\prime;p^\prime)$ is given by valid diagrams on the annulus (up to isotopy and local relations)
	\begin{equation}
		\begin{tikzpicture}[scale=1.2,baseline=(current bounding box.center)]
		\draw (0,0) circle (0.5cm);
		\draw (0,0) circle (1.6cm);
		\draw[brown,line width=0.4mm] (-60:0.5) -- (-60:1.6);
		\draw[violet,line width=0.4mm] (-120:0.5) -- (-120:1.6);
		\draw ([shift=(-60:1cm)]0,0) arc (-60:90:1cm);
		\draw ([shift=(-120:1.15cm)]0,0) arc (-120:-60:1.15cm);
		\draw ([shift=(240:0.85cm)]0,0) arc (240:90:0.85cm);
		\draw[nicegreen,line width=0.4mm] (90:0.5) -- (90:1.6);
		\node at (90:.3cm){$p$};\node at (90:1.8cm){$p'$};
		\node at (-60:.3cm){$n$};\node at (-60:1.8cm){$n'$};
		\node at (-120:.3cm){$m$};\node at (-120:1.8cm){$m'$};
		\node at (-1,0.3) {$a$};
		\node at (1.15,0.3) {$c$};
		\node at (0,-1.3) {$b$};
		\end{tikzpicture}\ .\label{eqn:basisann}
	\end{equation}
	For two diagrams $X,Y$ on the annulus, composition $Y\circ X$ corresponds to drawing the diagram $Y$ outside the diagram $X$ if the outer labels of $X$ match the inner labels of $Y$.
	The $F$-symbols of the fusion categories to reduce the composite diagram to a sum over diagrams of the form \eqref{eqn:basisann}.
	
	$N$-string annular categories can be defined analogously. The 1-string annular category, with the string labeled by $\mathcal{C}$ itself, coincides with the \emph{tube algebra} \cite{ocneanu}. In the case that the single string is labeled by an invertible bimodule, this coincides with the \emph{dube algebra} \cite{WBV17}. 
\end{definition}

\section{A $\Vec(\Z/2\Z)$ spin chain}
\label{Ising}

The chain model we want to study has the following structure: Given a fusion category $\mathcal{C}$, we take one distinguished object $X\in\mathcal{C}$ and study pairwise interaction within a chain of $N$ of these objects. To define a Hilbert space for this system, consider the fusion tree of the fusion process of $N$ objects of type $X$ as depicted below, which is a vector in the Hilbert space $C(X,X^{\otimes{N+1}})$.

\begin{figure}[H]
	\centering
	\begin{tikzpicture}[baseline=(current bounding box.center)]
		\draw (-0.25,0) -- (8.25,0);
		\draw (0.5,0) to node[right, pos=0.975] {$X$} (0.5,1);
		\draw (1.5,0) to node[right, pos=0.975] {$X$} (1.5,1);
		\draw (2.5,0) to node[right, pos=0.975] {$X$} (2.5,1);
		\draw (3.5,0) to node[right, pos=0.975] {$X$} (3.5,1);
		\draw (4.5,0) to node[right, pos=0.975] {$X$} (4.5,1);
		\draw (5.5,0) to node[right, pos=0.975] {$X$} (5.5,1);
		\draw (6.5,0) to node[right, pos=0.975] {$X$} (6.5,1);
		\draw (7.5,0) to node[right, pos=0.975] {$X$} (7.5,1);
		\node at (-0.1,-0.3) {$X$};
		\node at (1,-0.3) {$x_1$};
		\node at (2,-0.3) {$x_2$};
		\node at (3,-0.3) {$\dots$};
		\node at (4,-0.3) {};
		\node at (5,-0.3) {};
		\node at (6,-0.3) {};
		\node at (7,-0.3) {$x_{N-1}$};
		\node at (8.1,-0.3) {$X$};
	\end{tikzpicture}.
\end{figure}
\noindent
According to the fusion rules of the category that is considered, only specific combinations of the $x_1,x_2,\dots$ are allowed. The basis of this Hilbert space then corresponds to all admissible labelings $\lvert x_1,x_2,\dots,x_{N-1}\rangle$ of the links with $x_i\in L_\mathcal{C}$. Note that in the figure above, the boundary labels are fixed to $X$, but this is not mandatory. One can also consider these labels part of the labels that define the basis.

The interaction we want to study is a pairwise interaction where two neighboring particles favor to fuse to the vacuum $1$, i.e.\ we assign an energy gain to this fusion channel, e.g.\ $-1$. The local Hamiltonian that acts on sites $i$ and $i+1$ is then given by
	\begin{equation}\label{eq:Ham}
	h_i=-\frac{1}{\dim(X)}
	\begin{tikzpicture}[scale=0.5,baseline=(current bounding box.center)]
	\draw[] (0,0) to node [left,pos=0.8] {$X$} (-0.7,0.9);
	\draw[] (0,0) to node [right,pos=0.8] {$X$} (0.7,0.9);
	\draw (0,0) to node[right] {$1$} (0,-1);
	\draw[] (0,-1) to node [left,pos=0.8] {$X$} (-0.7,-1.9);
	\draw[] (0,-1) to node [right,pos=0.8] {$X$} (0.7,-1.9);
	\node at (0.8,-2.5) {$i+1$};
	\node at (-0.8,-2.5) {$i$};
	\end{tikzpicture}.
	\end{equation}
The complete Hamiltonian is then given by summing over all sites of the chain, $H=\sum_i h_i$.

In the example we want to study, $\Vec(\Z/2\Z)$, the objects are either $0$ or $1$ and the fusion is given by addition $\mathrm{mod}\ 2$. Note that here, the vacuum is denoted $0$. Two neighboring particles interact according to the fusion rules of $\Vec(\Z/2\Z)$, i.e. two identical particles fuse to $0$ and two different particles fuse to $1$. In order to define the Hilbert space for a fixed particle we can use fusion trees. These are chains of the following form:
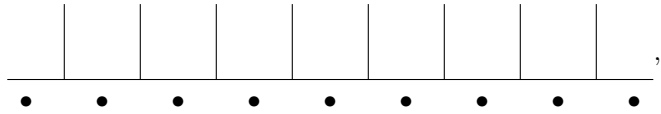
\begin{figure}[H]
	\begin{tikzpicture}[baseline=(current bounding box.center)]
	\draw (-0.25,0) -- (8.25,0);
	\draw (0.5,0) -- (0.5,1);
	\draw (1.5,0) -- (1.5,1);
	\draw (2.5,0) -- (2.5,1);
	\draw (3.5,0) -- (3.5,1);
	\draw (4.5,0) -- (4.5,1);
	\draw (5.5,0) -- (5.5,1);
	\draw (6.5,0) -- (6.5,1);
	\draw (7.5,0) -- (7.5,1);
	\node at (0,-0.3) {$\bullet$};
	\node at (1,-0.3) {$\bullet$};
	\node at (2,-0.3) {$\bullet$};
	\node at (3,-0.3) {$\bullet$};
	\node at (4,-0.3) {$\bullet$};
	\node at (5,-0.3) {$\bullet$};
	\node at (6,-0.3) {$\bullet$};
	\node at (7,-0.3) {$\bullet$};
	\node at (8,-0.3) {$\bullet$};
	\end{tikzpicture},
\end{figure}
\noindent
where the vertical lines are fixed and the bullets can either be $0$ or $1$, hence each bullet represents the space $\mathbb{C}^2$. The basis for the Hilbert space then corresponds to all possible labelings of the horizontal links which are allowed according to the fusion rules.
If we, for instance, only fuse $1$-particles to the chain, the value of the bullets is fixed by the fusion rules: When we fix the boundary labels (e.g., to $1$), the only possible labeling is
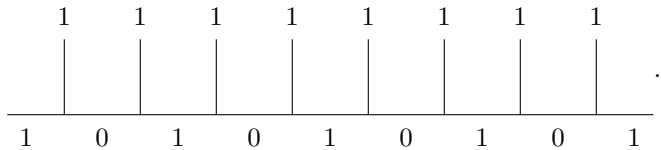
\begin{figure}[H]
	\begin{tikzpicture}[baseline=(current bounding box.center)]
	\draw (-0.25,0) -- (8.25,0);
	\draw (0.5,0) -- (0.5,1);
	\draw (1.5,0) -- (1.5,1);
	\draw (2.5,0) -- (2.5,1);
	\draw (3.5,0) -- (3.5,1);
	\draw (4.5,0) -- (4.5,1);
	\draw (5.5,0) -- (5.5,1);
	\draw (6.5,0) -- (6.5,1);
	\draw (7.5,0) -- (7.5,1);
	\node at (0.5,1.3) {$1$};
	\node at (1.5,1.3) {$1$};
	\node at (2.5,1.3) {$1$};
	\node at (3.5,1.3) {$1$};
	\node at (4.5,1.3) {$1$};
	\node at (5.5,1.3) {$1$};
	\node at (6.5,1.3) {$1$};
	\node at (7.5,1.3) {$1$};
	\node at (0,-0.3) {$1$};
	\node at (1,-0.3) {$0$};
	\node at (2,-0.3) {$1$};
	\node at (3,-0.3) {$0$};
	\node at (4,-0.3) {$1$};
	\node at (5,-0.3) {$0$};
	\node at (6,-0.3) {$1$};
	\node at (7,-0.3) {$0$};
	\node at (8,-0.3) {$1$};
	\end{tikzpicture}.
\end{figure}
\noindent
Hence, we have a unique ground state and the only vertices occurring in this case are
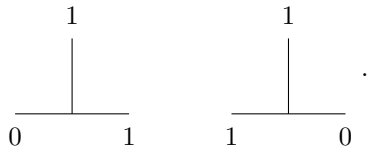
\begin{figure}[H]	
	\begin{tikzpicture}[baseline=(current bounding box.center)]
	\draw (0,0) -- (1.5,0);
	\draw (0.75,0) -- (0.75,1);
	\node at (0,-0.3) {$0$};
	\node at (1.5,-0.3) {$1$};
	\node at (0.75,1.3) {$1$};
	\end{tikzpicture}
	\hspace{20pt}
	\begin{tikzpicture}[baseline=(current bounding box.center)]
	\draw (0,0) -- (1.5,0);
	\draw (0.75,0) -- (0.75,1);
	\node at (0,-0.3) {$1$};
	\node at (1.5,-0.3) {$0$};
	\node at (0.75,1.3) {$1$};
	\end{tikzpicture}.
\end{figure}
\noindent
Analogously, if we only allow $0$s to fuse to the chain, and the outer labels are fixed to $1$, the only possible labeling is 
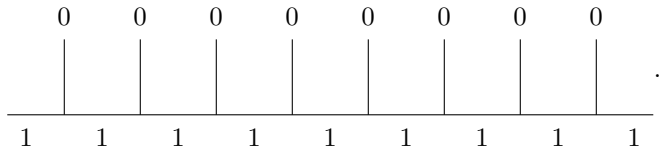
\begin{figure}[H]
	\begin{tikzpicture}[baseline=(current bounding box.center)]
	\draw (-0.25,0) -- (8.25,0);
	\draw (0.5,0) -- (0.5,1);
	\draw (1.5,0) -- (1.5,1);
	\draw (2.5,0) -- (2.5,1);
	\draw (3.5,0) -- (3.5,1);
	\draw (4.5,0) -- (4.5,1);
	\draw (5.5,0) -- (5.5,1);
	\draw (6.5,0) -- (6.5,1);
	\draw (7.5,0) -- (7.5,1);
	\node at (0.5,1.3) {$0$};
	\node at (1.5,1.3) {$0$};
	\node at (2.5,1.3) {$0$};
	\node at (3.5,1.3) {$0$};
	\node at (4.5,1.3) {$0$};
	\node at (5.5,1.3) {$0$};
	\node at (6.5,1.3) {$0$};
	\node at (7.5,1.3) {$0$};
	\node at (0,-0.3) {$1$};
	\node at (1,-0.3) {$1$};
	\node at (2,-0.3) {$1$};
	\node at (3,-0.3) {$1$};
	\node at (4,-0.3) {$1$};
	\node at (5,-0.3) {$1$};
	\node at (6,-0.3) {$1$};
	\node at (7,-0.3) {$1$};
	\node at (8,-0.3) {$1$};
	\end{tikzpicture}.
\end{figure}
\noindent
If we had fixed the outer labels to $0$, all bullets would have to be $0$. Hence, the only vertices occurring in these cases are
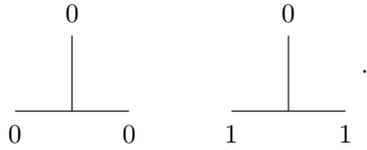
\begin{figure}[H]	
	\begin{tikzpicture}[baseline=(current bounding box.center)]
	\draw (0,0) -- (1.5,0);
	\draw (0.75,0) -- (0.75,1);
	\node at (0,-0.3) {$0$};
	\node at (1.5,-0.3) {$0$};
	\node at (0.75,1.3) {$0$};
	\end{tikzpicture}
	\hspace{20pt}
	\begin{tikzpicture}[baseline=(current bounding box.center)]
	\draw (0,0) -- (1.5,0);
	\draw (0.75,0) -- (0.75,1);
	\node at (0,-0.3) {$1$};
	\node at (1.5,-0.3) {$1$};
	\node at (0.75,1.3) {$0$};
	\end{tikzpicture}.
\end{figure}
\noindent
This means that in the case of open boundary conditions, the chain has a \emph{unique ground state} once we fix the outer labels of the chain. The situation is different for periodic boundary conditions: since the only requirement here is that the label at site $n+1$ has to equal the label at site $1$, there are always two possibilities of labeling the chain. For instance, in the case of only fusing $1$s to the chain, the two possibilities are
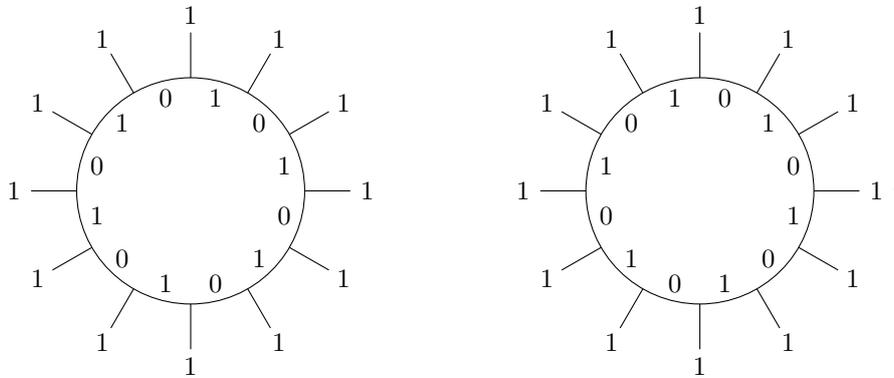
\begin{figure}[H]
	\begin{tikzpicture}[scale=1.5,baseline=(current bounding box.center)]
	\def\Radius{1cm}
	\draw (0,0) circle[radius=\Radius];
	\draw
	\foreach \a in {0, 30, ..., 330} {
		(\a:\Radius) -- (\a:1.4)
	};
	\foreach \a in {0, 30, ..., 330} {
		\node at (\a:1.55) {$1$};
	};
	\foreach \a in {45, 105, ..., 375} {
		\node at (\a:0.85) {$0$};
	};
	\foreach \a in {15, 75, ..., 315} {
		\node at (\a:0.85) {$1$};
	};
	\end{tikzpicture}
	\hspace{40pt}
	\begin{tikzpicture}[scale=1.5,baseline=(current bounding box.center)]
	\def\Radius{1cm}
	\draw (0,0) circle[radius=\Radius];
	\draw
	\foreach \a in {0, 30, ..., 330} {
		(\a:\Radius) -- (\a:1.4)
	};
	\foreach \a in {0, 30, ..., 330} {
		\node at (\a:1.55) {$1$};
	};
	\foreach \a in {45, 105, ..., 375} {
		\node at (\a:0.85) {$1$};
	};
	\foreach \a in {15, 75, ..., 315} {
		\node at (\a:0.85) {$0$};
	};
	\end{tikzpicture}.
\end{figure}
\noindent
In case of fusing $0$ to the chain, we can either allow only $0$s or only $1$ as labels of the chain. Therefore, the chain with periodic boundary conditions has two ground states, hence it represents a qubit. Note that all vertices that have occurred so far are defined within the fusion category.

We now introduce defects to this model, indicated by red lines that fuse to the chain, e.g.\ a chain with one defect would be
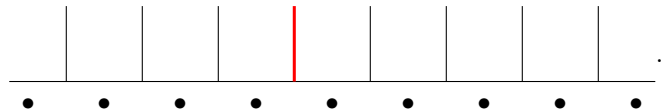
\begin{figure}[H]
	\begin{tikzpicture}[baseline=(current bounding box.center)]
	\draw (-0.25,0) -- (8.25,0);
	\draw (0.5,0) -- (0.5,1);
	\draw (1.5,0) -- (1.5,1);
	\draw (2.5,0) -- (2.5,1);
	\draw[red,line width=0.4mm] (3.5,0) -- (3.5,1);
	\draw (4.5,0) -- (4.5,1);
	\draw (5.5,0) -- (5.5,1);
	\draw (6.5,0) -- (6.5,1);
	\draw (7.5,0) -- (7.5,1);
	\node at (0,-0.3) {$\bullet$};
	\node at (1,-0.3) {$\bullet$};
	\node at (2,-0.3) {$\bullet$};
	\node at (3,-0.3) {$\bullet$};
	\node at (4,-0.3) {$\bullet$};
	\node at (5,-0.3) {$\bullet$};
	\node at (6,-0.3) {$\bullet$};
	\node at (7,-0.3) {$\bullet$};
	\node at (8,-0.3) {$\bullet$};
	\end{tikzpicture}.
\end{figure}
\noindent
Since the chain is represented by a category, $\Vec(\Z/2\Z)$ in our example, the defect is a $\Vec(\Z/2\Z)$-$\Vec(\Z/2\Z)$ bimodule, i.e.\ we are fusing an object from the bimodule to the chain. Here, we  use the bimodule $F_1$ (see Subsection \ref{sec:VecZp-bimodules}) to introduce defects to the $\Vec(\Z/2\Z)$-chain. $F_1$ has only one object, so in general we omit writing labels for the bimodule object, but indicate by a red line when the object is from the bimodule. When it is useful to indicate the label, we denote it $*$.

The occurrence of one defect in a chain does not change the ground state of the system; The labels on the chain are still determined by the choice of labels on the boundary. This is different if we have more than one defect in the chain. For instance, for two defects the chain is
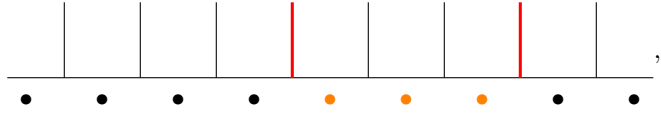
\begin{figure}[H]
	\begin{tikzpicture}[baseline=(current bounding box.center)]
	\draw (-0.25,0) -- (8.25,0);
	\draw (0.5,0) -- (0.5,1);
	\draw (1.5,0) -- (1.5,1);
	\draw (2.5,0) -- (2.5,1);
	\draw[red,line width=0.4mm] (3.5,0) -- (3.5,1);
	\draw (4.5,0) -- (4.5,1);
	\draw (5.5,0) -- (5.5,1);
	\draw[red,line width=0.4mm] (6.5,0) -- (6.5,1);
	\draw (7.5,0) -- (7.5,1);
	\node at (0,-0.3) {$\bullet$};
	\node at (1,-0.3) {$\bullet$};
	\node at (2,-0.3) {$\bullet$};
	\node at (3,-0.3) {$\bullet$};
	\node at (4,-0.3) {{\color{orange}$\bullet$}};
	\node at (5,-0.3) {{\color{orange}$\bullet$}};
	\node at (6,-0.3) {{\color{orange}$\bullet$}};
	\node at (7,-0.3) {$\bullet$};
	\node at (8,-0.3) {$\bullet$};
	\end{tikzpicture},
\end{figure}
\noindent
where the color of the orange bullets is not determined by the labels on the boundary of the chain. Hence, we get an additional qubit. In general, the number of additional qubits in the chain is $\#\mathrm{defects}-1$. This can be done analogously for the chain with periodic boundary conditions, the only difference is that in this case, we already have one qubit when there is no defect.

We could also allow the bimodule object to live on the horizontal lines of the chain. In particular,
We now want to build a chain out of $\Vec(\Z/2\Z)$ and the bimodule $F_1$ in the following way: consider the chain
\begin{figure}[H]
	\begin{tikzpicture}[baseline=(current bounding box.center)]
	\draw[blue,line width=0.4mm] (-0.25,0) -- (8.25,0);
	\draw[red,line width=0.4mm] (0.5,0) -- (0.5,1);
	\draw[red,line width=0.4mm] (1.5,0) -- (1.5,1);
	\draw[red,line width=0.4mm] (2.5,0) -- (2.5,1);
	\draw[red,line width=0.4mm] (3.5,0) -- (3.5,1);
	\draw[red,line width=0.4mm] (4.5,0) -- (4.5,1);
	\draw[red,line width=0.4mm] (5.5,0) -- (5.5,1);
	\draw[red,line width=0.4mm] (6.5,0) -- (6.5,1);
	\draw[red,line width=0.4mm] (7.5,0) -- (7.5,1);
	\end{tikzpicture},
\end{figure}
\noindent
where
\begin{equation}\label{eq:configs}
\begin{tikzpicture}[scale=1,baseline=(current bounding box.center)]
\draw[blue,line width=0.4mm] (0,0) -- (1,0);
\end{tikzpicture}=
\begin{tikzpicture}[scale=1,baseline=(current bounding box.center)]
\draw[black] (0,0) -- (1,0);
\end{tikzpicture}\oplus
\begin{tikzpicture}[scale=1,baseline=(current bounding box.center)]
\draw[red,line width=0.4mm] (0,0) -- (1,0);
\end{tikzpicture},
\end{equation}
which means that it is either an object from the category ($0$ or $1$) or an object from the bimodule (which can only be $*$), i.e., $\mathbb{C}^2\oplus\mathbb{C}\cong\mathbb{C}^3$. Valid configurations then look like
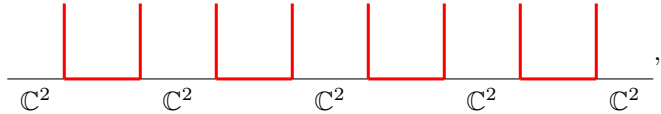
\begin{figure}[H]
	\begin{tikzpicture}[baseline=(current bounding box.center)]
	\draw[black] (-0.25,0) to node [below] {$\mathbb{C}^2$} (0.5,0);
	\draw[red,line width=0.4mm] (0.5,0) -- (0.5,1);
	\draw[red,line width=0.4mm] (0.5,0) -- (1.5,0);
	\draw[red,line width=0.4mm] (1.5,0) -- (1.5,1);
	\draw[black] (1.5,0) to node [below] {$\mathbb{C}^2$} (2.5,0);
	\draw[red,line width=0.4mm] (2.5,0) -- (2.5,1);
	\draw[red,line width=0.4mm] (2.5,0) -- (3.5,0);
	\draw[red,line width=0.4mm] (3.5,0) -- (3.5,1);
	\draw[black] (3.5,0) to node [below] {$\mathbb{C}^2$} (4.5,0);
	\draw[red,line width=0.4mm] (4.5,0) -- (4.5,1);
	\draw[red,line width=0.4mm] (4.5,0) -- (5.5,0);
	\draw[red,line width=0.4mm] (5.5,0) -- (5.5,1);
	\draw[black] (5.5,0) to node [below] {$\mathbb{C}^2$} (6.5,0);
	\draw[red,line width=0.4mm] (6.5,0) -- (6.5,1);
	\draw[red,line width=0.4mm] (6.5,0) -- (7.5,0);
	\draw[red,line width=0.4mm] (7.5,0) -- (7.5,1);
	\draw[black] (7.5,0) to node [below] {$\mathbb{C}^2$} (8.25,0);
	\end{tikzpicture},
\end{figure}
\noindent
so we have a non-trivial Hilbert space. Our goal is to construct a local Hamiltonian of the form \ref{eq:Ham} for this chain. For our model this means that it is energetically favored to fuse to the $0$ object of $\Vec(\Z/2\Z)$. Hence, the full Hamiltonian is of the form
\begin{equation}
H=-\sum \frac{1}{\sqrt{2}}
\begin{tikzpicture}[scale=0.5,baseline=(current bounding box.center)]
\draw[red,line width=0.4mm] (0,0) -- (-0.7,0.9);
\draw[red,line width=0.4mm] (0,0) -- (0.7,0.9);
\draw (0,0) to node[right] {$0$} (0,-1);
\draw[red,line width=0.4mm] (0,-1) -- (-0.7,-1.9);
\draw[red,line width=0.4mm] (0,-1) -- (0.7,-1.9);
\end{tikzpicture},\label{VecZ2Hamiltonian}
\end{equation}
where the sum goes over all sites of the chain and the normalization factor results from the fact that $\dim(*)=\sqrt{2}$. This Hamiltonian involves the vertex
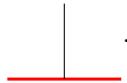
\begin{figure}[H]	
	\begin{tikzpicture}[baseline=(current bounding box.center)]
	\draw[red,line width=0.4mm] (0,0) -- (1.5,0);
	\draw[black] (0.75,0) -- (0.75,1);
	\end{tikzpicture}.
\end{figure}
\noindent
Since this vertex is neither defined in the category $\Vec(\Z/2\Z)$ nor in the bimodule $F_1$, we need to construct it. More precisely, for each choice of labels we need a basis for the corresponding morphism space. 
We do the construction of the required vertex step-by-step using ideas from tube algebras.
\newpage
\subsection{Computation of the bimodule vertex} 

\subsection*{Step 1: Compute isomorphism classes of objects and pick a representative.} In general, what we aim for are representations of the annular category of the form
	
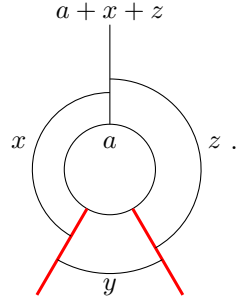
\begin{figure}[H]
		\begin{tikzpicture}[scale=1.2,baseline=(current bounding box.center)]
			\draw (0,0) circle (0.5cm);
			\draw[red,line width=0.4mm]
			\foreach \a in {-60, -120} {
				(\a:0.5) -- (\a:1.6)
			};
			\draw ([shift=(-60:1cm)]0,0) arc (-60:90:1cm);
			\draw ([shift=(-120:1.15cm)]0,0) arc (-120:-60:1.15cm);
			\draw ([shift=(240:0.85cm)]0,0) arc (240:90:0.85cm);
			\draw[] (90:0.5) -- (90:1.6);
			\node at (90:.3cm){$a$};\node at (90:1.75cm){$a+x+z$};
			\node at (-1,0.3) {$x$};
			\node at (1.15,0.3) {$z$};
			\node at (0,-1.3) {$y$};
		\end{tikzpicture}.
	\end{figure}
\noindent
According to the figure above, $(*,*,a)\cong(*,*,a+x+z)$ for all possible labels $x,y,z$, so we pick $(*,*,0)$ as a representative.

\subsection*{Step 2: Find primitive idempotents.} To find a representation of the annular category, we need to compute the primitive idempotents, i.e. we need to find morphisms that map $(*,*,0)$ to $(*,*,0)$ which square to themselves and are orthogonal to each other. Additionally, since we want \emph{primitive} idempotents, we need to make sure that they cannot be written as the sum of idempotents.
Candidates for idempotents are the morphism where $x+z=0$:
	\begin{equation}
		T_{0,0}=
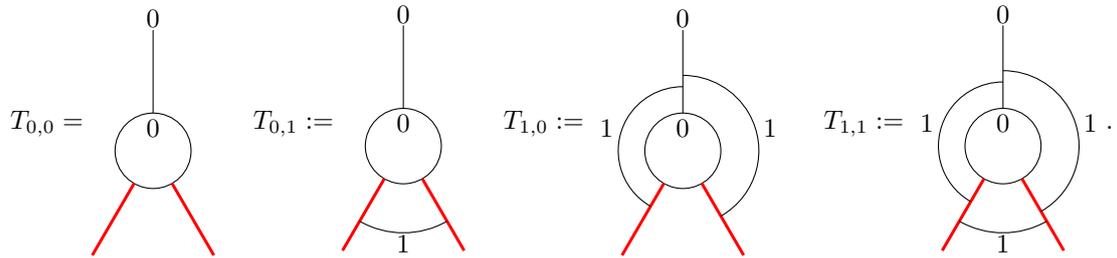
		\begin{tikzpicture}[scale=1,baseline=(current bounding box.center)]
			\draw (0,0) circle (0.5cm);
			\draw[red,line width=0.4mm]
			\foreach \a in {-60, -120} {
				(\a:0.5) -- (\a:1.6)
			};
			\draw[] (90:0.5) -- (90:1.6);
			\node at (90:.3cm){$0$};\node at (90:1.75cm){$0$};
		\end{tikzpicture}
		\hspace{5mm}
		T_{0,1}:=
		\begin{tikzpicture}[scale=1,baseline=(current bounding box.center)]
		\draw (0,0) circle (0.5cm);
		\draw[red,line width=0.4mm]
		\foreach \a in {-60, -120} {
			(\a:0.5) -- (\a:1.6)
		};
		\draw ([shift=(-120:1.15cm)]0,0) arc (-120:-60:1.15cm);
		\draw[] (90:0.5) -- (90:1.6);
		\node at (90:.3cm){$0$};\node at (90:1.75cm){$0$};
		\node at (0,-1.3) {$1$};
		\end{tikzpicture}
		\hspace{5mm}
		T_{1,0}:=
		\begin{tikzpicture}[scale=1,baseline=(current bounding box.center)]
		\draw (0,0) circle (0.5cm);
		\draw[red,line width=0.4mm]
		\foreach \a in {-60, -120} {
			(\a:0.5) -- (\a:1.6)
		};
		\draw ([shift=(-60:1cm)]0,0) arc (-60:90:1cm);
		\draw ([shift=(240:0.85cm)]0,0) arc (240:90:0.85cm);
		\draw[] (90:0.5) -- (90:1.6);
		\node at (90:.3cm){$0$};\node at (90:1.75cm){$0$};
		\node at (-1,0.3) {$1$};
		\node at (1.15,0.3) {$1$};
		\end{tikzpicture}
		\hspace{5mm}
		T_{1,1}:=
		\begin{tikzpicture}[scale=1,baseline=(current bounding box.center)]
		\draw (0,0) circle (0.5cm);
		\draw[red,line width=0.4mm]
		\foreach \a in {-60, -120} {
			(\a:0.5) -- (\a:1.6)
		};
		\draw ([shift=(-60:1cm)]0,0) arc (-60:90:1cm);
		\draw ([shift=(-120:1.15cm)]0,0) arc (-120:-60:1.15cm);
		\draw ([shift=(240:0.85cm)]0,0) arc (240:90:0.85cm);
		\draw[] (90:0.5) -- (90:1.6);
		\node at (90:.3cm){$0$};\node at (90:1.75cm){$0$};
		\node at (-1,0.3) {$1$};
		\node at (1.15,0.3) {$1$};
		\node at (0,-1.3) {$1$};
		\end{tikzpicture}.
	\end{equation}
\noindent
The first morphism can be interpreted as the identity morphism and it obviously squares to itself. It is easy to see that the second and third diagrams square to the first. To square the final morphism, we need to do the following calculation:
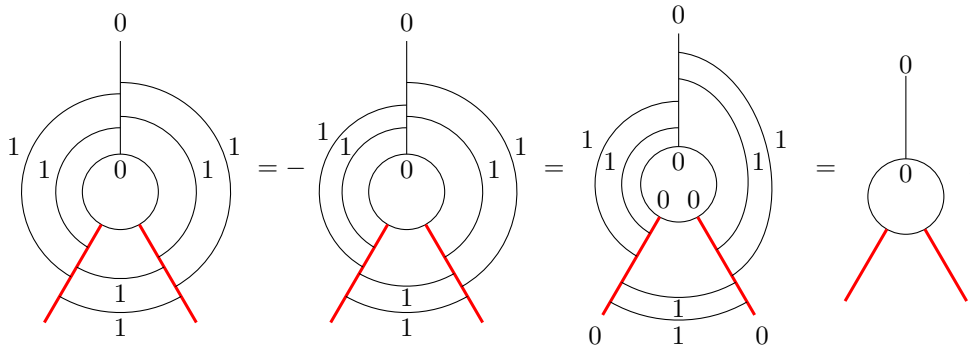
\vspace{5pt}
	\begin{equation*}
		\begin{tikzpicture}[scale=1,baseline=(current bounding box.center)]
			\draw (0,0) circle (0.5cm);
			\draw[red,line width=0.4mm]
			\foreach \a in {-60, -120} {
				(\a:0.5) -- (\a:2)
			};
			\draw[] (90:0.5) -- (90:2);
			\draw ([shift=(-60:1cm)]0,0) arc (-60:90:1cm);
			\draw ([shift=(-120:1.15cm)]0,0) arc (-120:-60:1.15cm);
			\draw ([shift=(240:0.85cm)]0,0) arc (240:90:0.85cm);
			\draw ([shift=(240:1.3cm)]0,0) arc (240:90:1.3cm);
			\draw ([shift=(-60:1.45cm)]0,0) arc (-60:90:1.45cm);
			\draw ([shift=(-120:1.6cm)]0,0) arc (-120:-60:1.6cm);
			\node at (90:.3cm){$0$};\node at (90:2.25cm){$0$};
			\node at (-1,0.3) {$1$};
			\node at (1.15,0.3) {$1$};
			\node at (0,-1.35) {$1$};
			\node at (-1.4,0.6) {$1$};
			\node at (1.5,0.6) {$1$};
			\node at (0,-1.8) {$1$};
		\end{tikzpicture}
		=-\begin{tikzpicture}[scale=1,baseline=(current bounding box.center)]
			\draw (0,0) circle (0.5cm);
			\draw[red,line width=0.4mm]
			\foreach \a in {-60, -120} {
				(\a:0.5) -- (\a:2)
			};
			\draw[] (90:0.5) -- (90:2);
			\draw ([shift=(-60:1cm)]0,0) arc (-60:90:1cm);
			\draw ([shift=(-120:1.25cm)]0,0) arc (-120:-60:1.25cm);
			\draw ([shift=(240:0.85cm)]0,0) arc (240:90:0.85cm);
			\draw ([shift=(240:1.15cm)]0,0) arc (240:90:1.15cm);
			\draw ([shift=(-60:1.45cm)]0,0) arc (-60:90:1.45cm);
			\draw ([shift=(-120:1.6cm)]0,0) arc (-120:-60:1.6cm);
			\node at (90:.3cm){$0$};\node at (90:2.25cm){$0$};
			\node at (-1.1,0.8) {$1$};
			\node at (1.15,0.3) {$1$};
			\node at (0,-1.4) {$1$};
			\node at (-0.8,0.6) {$1$};
			\node at (1.5,0.6) {$1$};
			\node at (0,-1.8) {$1$};
		\end{tikzpicture}
		=\begin{tikzpicture}[scale=1,baseline=(current bounding box.center)]
			\draw (0,0) circle (0.5cm);
			\draw[red,line width=0.4mm]
			\foreach \a in {-60, -120} {
				(\a:0.5) -- (\a:2)
			};
			\draw[] (90:0.5) -- (90:2);
			\draw ([shift=(240:0.75cm)]0,0) arc (240:90:0.75cm);
			\draw ([shift=(-60:1cm)]0,0) to [bend right=70] ([shift=(90:1.4cm)]0,0);
			\draw ([shift=(-120:1.5cm)]0,0) arc (-120:-60:1.5cm);
			\draw ([shift=(240:1.1cm)]0,0) arc (240:90:1.1cm);
			\draw ([shift=(-60:1.4cm)]0,0) to [bend right=70] ([shift=(90:1.75cm)]0,0);
			\draw ([shift=(-120:1.8cm)]0,0) arc (-120:-60:1.8cm);
			\node at (90:.3cm){$0$};\node at (90:2.25cm){$0$};
			\node at (0.2,-0.2) {$0$};
			\node at (-0.2,-0.2) {$0$};
			\node at (-1.1,-2) {$0$};
			\node at (1.1,-2) {$0$};
			\node at (-0.9,0.3) {$1$};
			\node at (1.05,0.3) {$1$};
			\node at (0,-1.65) {$1$};
			\node at (-1.2,0.6) {$1$};
			\node at (1.35,0.6) {$1$};
			\node at (0,-2) {$1$};
		\end{tikzpicture}
		=\begin{tikzpicture}[scale=1,baseline=(current bounding box.center)]
			\draw (0,0) circle (0.5cm);
			\draw[red,line width=0.4mm]
			\foreach \a in {-60, -120} {
				(\a:0.5) -- (\a:1.6)
			};
			\draw[] (90:0.5) -- (90:1.6);
			\node at (90:.3cm){$0$};\node at (90:1.75cm){$0$};
		\end{tikzpicture}\vspace{5pt}.
	\end{equation*}
Hence, the second candidate does not square to itself but to the first candidate. 
Similar computations show that $T_{a,b}T_{c,d}=T_{a+c,b+d}$. 
However, since the primitive idempotents form an algebra, we can also consider linear combinations of the candidates. Also, because we have four candidates, we know that the algebra is $4$-dimensional. To find out the primitive idempotents from the candidates, it is convenient to find a representation of them in terms of matrices, i.e.\ we need matrices that multiply in the same way as the candidates. As mentioned above, the first candidate is the identity, hence the corresponding matrix is
	\begin{equation}
		M_{0,0}=\begin{pmatrix}
			1 & 0 & 0 & 0\\
			0 & 1 & 0 & 0\\
			0 & 0 & 1 & 0\\
			0 & 0 & 0 & 1\\
		\end{pmatrix}.
	\end{equation}
From this representation it is clear that this candidate is an idempotent, but not a primitive one: $M_{0,0}$ can be written as
	\begin{equation}
	\label{eq_M0}
		M_{0,0}=
		\begin{pmatrix}
		1 & 0 & 0 & 0\\
		0 & 0 & 0 & 0\\
		0 & 0 & 0 & 0\\
		0 & 0 & 0 & 0\\
		\end{pmatrix}
		+\begin{pmatrix}
		0 & 0 & 0 & 0\\
		0 & 1 & 0 & 0\\
		0 & 0 & 0 & 0\\
		0 & 0 & 0 & 0\\
		\end{pmatrix}
		+
		\begin{pmatrix}
		0 & 0 & 0 & 0\\
		0 & 0 & 0 & 0\\
		0 & 0 & 1 & 0\\
		0 & 0 & 0 & 0\\
		\end{pmatrix}
		+\begin{pmatrix}
		0 & 0 & 0 & 0\\
		0 & 0 & 0 & 0\\
		0 & 0 & 0 & 0\\
		0 & 0 & 0 & 1\\
		\end{pmatrix}
		.
	\end{equation}
The matrix representing the second candidate has to fulfill $M_{0,1}^2=M_{0,0}$, so a possible choice is
	\begin{equation}
		M_{0,1}=\begin{pmatrix}
			1 & 0 & 0 & 0\\
			0 & 1 & 0 & 0\\
			0 & 0 & -1 & 0\\
			0 & 0 & 0 & -1\\
		\end{pmatrix}.
	\end{equation}
	Note that this representation is not unique. Similar matrices can be found for the remaining morphisms.
From \eqref{eq_M0}, we get four candidates for primitive idempotents. They are indeed primitive idempotents in the algebra of $4$-dimensional matrices. To translate them back into annular diagrams, we just need to express them in terms of $M_{a,b}$, which yields
	\begin{align}
	P_{x,y}=\frac{1}{4}\sum_{a,b}(-1)^{ax+by}M_{a,b}
	\end{align}
which is diagrammatically
	\begin{equation}
		P_{x,y}=\frac{1}{4}\left(
			\begin{tikzpicture}[scale=1,baseline=(current bounding box.center)]
			\draw (0,0) circle (0.5cm);
			\draw[red,line width=0.4mm]
			\foreach \a in {-60, -120} {
				(\a:0.5) -- (\a:1.6)
			};
			\draw[] (90:0.5) -- (90:1.6);
			\node at (90:.3cm){$0$};\node at (90:1.75cm){$0$};
			\end{tikzpicture}
			+(-1)^x
			\begin{tikzpicture}[scale=1,baseline=(current bounding box.center)]
			\draw (0,0) circle (0.5cm);
			\draw[red,line width=0.4mm]
			\foreach \a in {-60, -120} {
				(\a:0.5) -- (\a:1.6)
			};
			\draw ([shift=(-120:1.15cm)]0,0) arc (-120:-60:1.15cm);
			\draw[] (90:0.5) -- (90:1.6);
			\node at (90:.3cm){$0$};\node at (90:1.75cm){$0$};
						\node at (0,-1.3) {$1$};
			\end{tikzpicture}
			+(-1)^y
			\begin{tikzpicture}[scale=1,baseline=(current bounding box.center)]
			\draw (0,0) circle (0.5cm);
			\draw[red,line width=0.4mm]
			\foreach \a in {-60, -120} {
				(\a:0.5) -- (\a:1.6)
			};
									\draw ([shift=(-60:1cm)]0,0) arc (-60:90:1cm);
									\draw ([shift=(240:0.85cm)]0,0) arc (240:90:0.85cm);
			\draw[] (90:0.5) -- (90:1.6);
			\node at (90:.3cm){$0$};\node at (90:1.75cm){$0$};
						\node at (-1,0.3) {$1$};
									\node at (1.15,0.3) {$1$};
			\end{tikzpicture}
			+(-1)^{x+y}
			\begin{tikzpicture}[scale=1,baseline=(current bounding box.center)]
			\draw (0,0) circle (0.5cm);
			\draw[red,line width=0.4mm]
			\foreach \a in {-60, -120} {
				(\a:0.5) -- (\a:1.6)
			};
									\draw ([shift=(-60:1cm)]0,0) arc (-60:90:1cm);
			\draw ([shift=(-120:1.15cm)]0,0) arc (-120:-60:1.15cm);
									\draw ([shift=(240:0.85cm)]0,0) arc (240:90:0.85cm);
			\draw[] (90:0.5) -- (90:1.6);
			\node at (90:.3cm){$0$};\node at (90:1.75cm){$0$};
			\node at (-1,0.3) {$1$};
			\node at (1.15,0.3) {$1$};
			\node at (0,-1.3) {$1$};
			\end{tikzpicture}
			\right).
	\end{equation}
	
\subsection*{Step 3: Check for isomorphism classes of primitive idempotents.} In general, it is possible that the primitive idempotents we just found are isomorphic to each other, which means that there are matrices within the algebra that we can multiply to $P_{0,0}$, for example, and get $P_{1,0}$. The following equation is an example for how this works:
	\begin{equation}
		\begin{pmatrix} 0 & 0 & 0 & 0\\ 1 & 0 & 0 & 0\\ 0 & 0 & 0 & 0\\ 0 & 0 & 0 & 0\\ \end{pmatrix}P_{0,0} \begin{pmatrix} 0 & 1 & 0 & 0\\ 0 & 0 & 0 & 0\\ 0 & 0 & 0 & 0\\ 0 & 0 & 0 & 0\\ \end{pmatrix}=P_{1,0}.\label{eqn:isoids}
	\end{equation}
However, the two matrices we multiply $P_{0,0}$ with are not elements of the matrix algebra (the matrices on the algebra only have entries on the diagonal). If they were elements of the algebra, they would form an isomorphism between $P_{0,0}$ and $P_{1,0}$, so we would pick either $P_{0,0}$ or $P_{1,0}$ as a representative.

This step can be much more complicated for algebras with bigger dimensions. However, there is a nice trick which helps to see how many isomorphism classes there are, which is the Artin-Wedderburn theorem (see, for example, \cite{Beachy1999}). It states that any semi-simple algebra can be decomposed into a direct sum of full matrix algebras, i.e.
	\begin{equation}
		\mathcal{M}\cong\bigoplus\mathcal{M}_d,
	\end{equation}
	where $\dim{\mathcal{M}_d}=d^2$. Within a full matrix algebra we can pick, for example, the primitive idempotent $\mathrm{diag}(1,0,\ldots)$ since the equivalent of the matrices in  \eqref{eqn:isoids} exist. We therefore only get one primitive idempotent for each full matrix algebra. In our case, the algebra is 4 dimensional. The possible decompositions are therefore 
	$\mathcal{M}\cong\mathbb{C}\oplus\mathbb{C}\oplus\mathbb{C}\oplus\mathbb{C}$ and $\mathcal{M}_2$. We found the first decomposition was correct so we get four primitive idempotents (as we showed above). 
	In case of a $5$-dimensional matrix algebra, the possible decompositions are
	\begin{equation}
		\mathcal{M}_{5-\mathrm{dim}}\cong\mathbb{C}\oplus\mathcal{M}_2(\mathbb{C}),
	\end{equation}
	or 
	\begin{equation}
		\mathcal{M}_{5-\mathrm{dim}}\cong5\mathbb{C},
	\end{equation}
so we get two or five primitive idempotents respectively.
	
\subsection*{Step 4: Build the full representation.} After we have found the primitive idempotents of the algebra, we can build the full representation. This is done by putting all possible annuli on the outside of the idempotents, hence finding all the basis vectors for this space, i.e.\ all possible vectors
	\begin{equation}
		\begin{tikzpicture}[scale=1,baseline=(current bounding box.center)]
		\draw (0,0) circle (0.5cm);
		\draw[red,line width=0.4mm]
		\foreach \a in {-60, -120} {
			(\a:0.5) -- (\a:1.6)
		};
		\draw ([shift=(-60:1cm)]0,0) arc (-60:90:1cm);
		\draw ([shift=(-120:1.15cm)]0,0) arc (-120:-60:1.15cm);
		\draw ([shift=(240:0.85cm)]0,0) arc (240:90:0.85cm);
		\draw[] (90:0.5) -- (90:1.6);
		\node at (90:.3cm){$0$};\node at (90:1.75cm){$\alpha+\gamma$};
		\node at (-1,0.3) {$\alpha$};
		\node at (1.15,0.3) {$\gamma$};
		\node at (0,-1.4) {$\beta$};
		\end{tikzpicture}.
	\end{equation}
In general, the basis vectors are determined by the choice of $\alpha,\beta$ and $\gamma$, therefore there are up to $2^3=8$ possible basis vectors for each representation. However, it is possible that some of these vectors are linearly dependent, which is the case in our example, as we will see. Putting the general annuli around the primitive idempotents $P_{x,y}$ yields
	\begin{align}
	\frac{1}{4}
	\sum_{a,b}(-1)^{ax+by}
	\begin{tikzpicture}[scale=1,baseline=(current bounding box.center)]
	\draw (0,0) circle (0.5cm);
	\draw[red,line width=0.4mm]
	\foreach \a in {-60, -120} {
		(\a:0.5) -- (\a:2)
	};
	\draw[] (90:0.5) -- (90:2);
	\draw ([shift=(-60:1cm)]0,0) arc (-60:90:1cm);
	\draw ([shift=(-120:1.15cm)]0,0) arc (-120:-60:1.15cm);
	\draw ([shift=(240:0.85cm)]0,0) arc (240:90:0.85cm);
	\draw ([shift=(240:1.3cm)]0,0) arc (240:90:1.3cm);
	\draw ([shift=(-60:1.45cm)]0,0) arc (-60:90:1.45cm);
	\draw ([shift=(-120:1.6cm)]0,0) arc (-120:-60:1.6cm);
	\node at (90:.3cm){$0$};\node at (90:2.25cm){$\alpha+\gamma$};
	\node at (-1,0.3) {$b$};
	\node at (1.15,0.3) {$b$};
	\node at (0,-1.35) {$a$};
	\node at (-1.4,0.6) {$\alpha$};
	\node at (1.5,0.6) {$\gamma$};
	\node at (0,-1.8) {$\beta$};
	\end{tikzpicture}
	&=
	\frac{1}{4}
	\sum_{a,b}(-1)^{a(x+\alpha+\gamma)+by}
	\begin{tikzpicture}[scale=1,baseline=(current bounding box.center)]
	\draw (0,0) circle (0.5cm);
	\draw[red,line width=0.4mm]
	\foreach \a in {-60, -120} {
		(\a:0.5) -- (\a:1.6)
	};
	\draw ([shift=(-60:1cm)]0,0) arc (-60:90:1cm);
	\draw ([shift=(-120:1.15cm)]0,0) arc (-120:-60:1.15cm);
	\draw ([shift=(240:0.85cm)]0,0) arc (240:90:0.85cm);
	\draw[] (90:0.5) -- (90:1.6);
	\node at (90:.3cm){$0$};\node at (90:1.75cm){$\alpha+\gamma$};
	\node[rotate=65] at (-1,0.3) {$\alpha+b$};
	\node[rotate=-65] at (1.15,0.3) {$\gamma+b$};
	\node at (0,-1.3) {$\beta+a$};
	\end{tikzpicture}\\
	&=
	\frac{(-1)^{(x+\alpha^\prime)\beta+y\gamma}}{4}
	\sum_{a^\prime,b^\prime}(-1)^{a^\prime(x+\alpha^\prime)+b^\prime y}
	\begin{tikzpicture}[scale=1,baseline=(current bounding box.center)]
	\draw (0,0) circle (0.5cm);
	\draw[red,line width=0.4mm]
	\foreach \a in {-60, -120} {
		(\a:0.5) -- (\a:1.6)
	};
	\draw ([shift=(-60:1cm)]0,0) arc (-60:90:1cm);
	\draw ([shift=(-120:1.15cm)]0,0) arc (-120:-60:1.15cm);
	\draw ([shift=(240:0.85cm)]0,0) arc (240:90:0.85cm);
	\draw[] (90:0.5) -- (90:1.6);
	\node at (90:.3cm){$0$};\node at (90:1.75cm){$\alpha^\prime$};
	\node[rotate=65] at (-1,0.3) {$\alpha^\prime+b^\prime$};
	\node at (1.15,0.3) {$b^\prime$};
	\node at (0,-1.3) {$a^\prime$};
	\end{tikzpicture}
	\end{align}
which is a vector in the morphism space
	\begin{equation}	
		\begin{tikzpicture}
			\draw[red,line width=0.4mm] (0,0) -- (1.5,0);
			\draw (0.75,0) -- (0.75,1) node [pos=1.25]{$\alpha^\prime$};
		\end{tikzpicture}.
	\end{equation}
We now have to find a basis for every one of those morphism spaces, i.e.\ for every choice of $\alpha^\prime$. For fixed representation (fixed, $x,y$), we find a unique vector up to a multiplicative scalar for each $\alpha^\prime$, so each morphism space is one dimensional. We define the basis to be
	\begin{equation}
		\begin{tikzpicture}[scale=1,baseline=(current bounding box.center)]
			\draw[red,line width=0.4mm] (0,-.25) -- (.75,0) -- (1.5,-.25);
			\draw (0.75,0) -- (0.75,1) node [pos=1.25] {$\alpha$};
			\node[above,right] at (.75,.15) {$(x,y)$};
		\end{tikzpicture}\equiv
		\frac{1}{4}
		\sum_{a,b}(-1)^{a(x+\alpha)+b y}
		\begin{tikzpicture}[scale=1,baseline=(current bounding box.center)]
		\draw (0,0) circle (0.5cm);
		\draw[red,line width=0.4mm]
		\foreach \a in {-60, -120} {
			(\a:0.5) -- (\a:1.6)
		};
		\draw ([shift=(-60:1cm)]0,0) arc (-60:90:1cm);
		\draw ([shift=(-120:1.15cm)]0,0) arc (-120:-60:1.15cm);
		\draw ([shift=(240:0.85cm)]0,0) arc (240:90:0.85cm);
		\draw[] (90:0.5) -- (90:1.6);
		\node at (90:.3cm){$0$};\node at (90:1.75cm){$\alpha$};
		\node[rotate=65] at (-1,0.3) {$\alpha+b$};
		\node at (1.15,0.3) {$b$};
		\node at (0,-1.3) {$a$};
		\end{tikzpicture}.
	\end{equation}
The `inflation trick' developed in \cite{BBJ18,BB19a,BB19b} picks out the representation $P_{0,0}$, and we work with that from here on. This means
	\begin{equation}
	\begin{tikzpicture}[scale=1,baseline=(current bounding box.center)]
	\draw[red,line width=0.4mm] (0,-.25) -- (.75,0) -- (1.5,-.25);
	\draw (0.75,0) -- (0.75,1) node [pos=1.25] {$\alpha$};
	\end{tikzpicture}:=
	\frac{1}{4}
	\sum_{a,b}(-1)^{a\alpha}
	\begin{tikzpicture}[scale=1,baseline=(current bounding box.center)]
	\draw (0,0) circle (0.5cm);
	\draw[red,line width=0.4mm]
	\foreach \a in {-60, -120} {
		(\a:0.5) -- (\a:1.6)
	};
	\draw ([shift=(-60:1cm)]0,0) arc (-60:90:1cm);
	\draw ([shift=(-120:1.15cm)]0,0) arc (-120:-60:1.15cm);
	\draw ([shift=(240:0.85cm)]0,0) arc (240:90:0.85cm);
	\draw[] (90:0.5) -- (90:1.6);
	\node at (90:.3cm){$0$};\node at (90:1.75cm){$\alpha$};
	\node[rotate=65] at (-1,0.3) {$\alpha+b$};
	\node at (1.15,0.3) {$b$};
	\node at (0,-1.3) {$a$};
	\end{tikzpicture}
	\end{equation}

\subsection*{Step 5: Find associator of the extended category.} After the vertex itself is defined, we want to compute the $F$-symbols related to the new object. 
Our goal is to define a Hamiltonian for the chain which will be of the form
	\begin{equation*}
		\begin{tikzpicture}[scale=1,baseline=(current bounding box.center)]
		\draw (.5,.25)--(.5,.75) node [pos=.5,left]{0};
		\draw [red, line width=0.25mm] (0,0) -- (.5,.25) -- (1,0);
		\draw [red, line width=0.25mm] (0,1) -- (.5,.75)--(1,1);
		\end{tikzpicture}
		=
		\alpha\ \begin{tikzpicture}[scale=1,baseline=(current bounding box.center)]
			\draw [red, line width=0.25mm] (0,0) -- (0,1);
			\draw [red, line width=0.25mm] (0.5,0) -- (0.5,1);
		\end{tikzpicture}
		+\beta\ \begin{tikzpicture}[scale=1,baseline=(current bounding box.center)]
			\draw (0,0.5) -- (0.5,0.5);
			\draw [red, line width=0.25mm] (0,0) -- (0,1);
			\draw [red, line width=0.25mm] (0.5,0) -- (0.5,1);			
		\end{tikzpicture}\ .
	\end{equation*}
To compute $\alpha$ and $\beta$ above, in addition to the action on the spin chain, we will need the full set of $F$-symbols.

From the category $\Vec(\Z/2\Z)$, and the bimodule associators we have
\begin{align}
\left(F_{abc}^{a+b+c}\right)_{a+b,b+c}&=1&&\text{From Example~\ref{example:vecG}: $F$=+1}\\
\left(F_{ab*}^*\right)_{a+b,*}&=1&&\text{From \eqref{eqn:L}: $L$=+1} \\
\left(F_{a*b}^*\right)_{*,*}&=(-1)^{ab}&&\text{From \eqref{eqn:F1}} \\
\left(F_{*ab}^*\right)_{*,a+b}&=1&&\text{From \eqref{eqn:R}: $R$=+1}.
\end{align}
\noindent
However, there are still some yet unknown $F$-symbols, namely
\begin{align}
	\left(F_{a**}^{a+b}\right)_{*,b}&=??\\
	\left(F_{*a*}^b\right)_{*,*}&=??\\
	\left(F_{**a}^{a+b}\right)_{b,*}&=??\\
	\left(F_{***}^*\right)_{a,b}&=??.
\end{align}
\noindent	
To compute those, we use the following normalization \cite{Bonderson,BSS08}:
	\begin{equation}
		\begin{tikzpicture}[scale=1.2,baseline=(current bounding box.center)]
			\draw (0,0) to node [left] {$a$} (0,1);
			\draw (0.5,0) to node [right] {$b$} (0.5,1);
		\end{tikzpicture}=\sum_c\ \sqrt{\frac{d_c}{d_a d_b}}\ 
		\begin{tikzpicture}[scale=1.2,baseline=(current bounding box.center)]
			\draw (0,0) to node [left] {$a$} (0.25,0.25);
			\draw (0.5,0) to node [right] {$b$} (0.25,0.25);
			\draw (0,1) to node [left] {$a$} (0.25,0.75);
			\draw (0.5,1) to node [right] {$b$} (0.25,0.75);
			\draw (0.25,0.25) to node [right] {$c$} (0.25,0.75);
		\end{tikzpicture}.
	\end{equation}
For the black strings in our diagrams, the sum has only one term and the coefficients are $d_0=1=d_1$, which yields the relation
	\begin{equation}
	\label{eq:completeness}
		\begin{tikzpicture}[scale=1.2,baseline=(current bounding box.center)]
			\draw (0,0) to node [left] {$a$} (0,1);
			\draw (0.5,0) to node [right] {$b$} (0.5,1);
		\end{tikzpicture}=
		\begin{tikzpicture}[scale=1.2,baseline=(current bounding box.center)]
			\draw (0,0) to node [left] {$a$} (0.25,0.25);
			\draw (0.5,0) to node [right] {$b$} (0.25,0.25);
			\draw (0,1) to node [left] {$a$} (0.25,0.75);
			\draw (0.5,1) to node [right] {$b$} (0.25,0.75);
			\draw (0.25,0.25) to node [right] {$a+b$} (0.25,0.75);
		\end{tikzpicture}.
	\end{equation}
From the fusion rules, $d_*=\sqrt{2}$, so
\begin{equation}
\label{eq:completeness2}
\begin{tikzpicture}[scale=1.2,baseline=(current bounding box.center)]
\draw[red,line width=0.4mm] (0,0) to (0,1);
\draw (0.5,0) to node [right] {$a$} (0.5,1);
\end{tikzpicture}=
\begin{tikzpicture}[scale=1.2,baseline=(current bounding box.center)]
\draw[red,line width=0.4mm] (0,0) to (0.25,0.25);
\draw (0.5,0) to node [right] {$a$} (0.25,0.25);
\draw[red,line width=0.4mm] (0,1) to (0.25,0.75);
\draw (0.5,1) to node [right] {$a$} (0.25,0.75);
\draw[red,line width=0.4mm] (0.25,0.25) to (0.25,0.75);
\end{tikzpicture}.
\end{equation}
The computation for $F_{**a}^{a+b}$ is particularly straightforward:
	\begin{align}
		\begin{tikzpicture}[scale=0.7,baseline=(current bounding box.center)]
			\draw[] (0,0) -- (0,1) node[pos=1,above]{$a+b$};
			\draw[] (0,0) -- (-0.707107,-0.707107) node [pos=.5,left] {$b$};
			\draw[red,line width=0.4mm] (-0.707107,-0.707107) -- (-1.41421,-1.41421);
			\draw[red,line width=0.4mm] (-0.707107,-0.707107) -- (0,-1.41421);
			\draw (0,0) -- (0.707107,-0.707107);
			\draw (0.707107,-0.707107) -- (1.41421,-1.41421);
			\node at (-1.41421,-1.7) {$*$};
			\node at (0,-1.7) {$*$};
			\node at (1.41421,-1.7) {$a$};
		\end{tikzpicture}&=
		\frac{1}{4}\sum_{x,y}(-1)^{bx}
		\begin{tikzpicture}[scale=.75,baseline=(current bounding box.center)]
			\draw (0,0) circle (0.5cm);
			\draw[red,line width=0.4mm]
			\foreach \a in {-60, -120} {
				(\a:0.5) -- (\a:1.6)
			};
			\draw[] (90:0.5) -- (90:1.6);
			\draw ([shift=(-60:1cm)]0,0) arc (-60:90:1cm);
			\draw ([shift=(-120:1.15cm)]0,0) arc (-120:-60:1.15cm);
			\draw ([shift=(240:0.85cm)]0,0) arc (240:90:0.85cm);
			\draw[] (0,1.6) -- (0.9,2.1) node[pos=0,left] {$b$};
			\draw[] (0.9,2.1) -- (0.9,2.6);
			\draw (0.9,2.1) -- (2.3,-1.35);
			\node[rotate=60] at (-1,0.3) {$b+y$};
			\node at (1.15,0.3) {$y$};
			\node at (2.3,-1.7) {$a$};
			\node at (0,-1.35) {$x$};
		\end{tikzpicture}
		=\frac{1}{4}\sum_{x,y}(-1)^{bx}
		\begin{tikzpicture}[scale=.75,baseline=(current bounding box.center)]
		\draw (0,0) circle (0.5cm);
		\draw[red,line width=0.4mm]
		\foreach \a in {-60, -120} {
			(\a:0.5) -- (\a:2.5)
		};
		\draw[] (90:0.5) -- (90:1.6);
		\draw ([shift=(-60:1cm)]0,0) arc (-60:90:1cm);
		\draw ([shift=(-120:1.15cm)]0,0) arc (-120:-60:1.15cm);
		\draw ([shift=(240:0.85cm)]0,0) arc (240:90:0.85cm);
		\draw[] (0,1.6) -- (0.9,2.1) node[pos=0,left] {$b$};
		\draw[] (0.9,2.1) -- (0.9,2.6);
		\draw (-60:1.6) -- (2.3,-2) node[pos=1,below] {$a$};
		\draw (-60:1.4) --(1.5,0) node[right]{$a$} -- (0.9,2.1);
		\node[rotate=60] at (-1,0.3) {$b+y$};
		\node at (1.15,0.3) {$y$};
		\node at (0,-1.35) {$x$};
		\end{tikzpicture}
		\\
		&=\frac{1}{4}\sum_{x,y}(-1)^{(a+b)x}
		\begin{tikzpicture}[scale=.75,baseline=(current bounding box.center)]
		\draw (0,0) circle (0.5cm);
		\draw[red,line width=0.4mm]
		\foreach \a in {-60, -120} {
			(\a:0.5) -- (\a:2.5)
		};
		\draw[] (90:0.5) -- (90:1.6);
		\draw ([shift=(-60:1cm)]0,0) arc (-60:90:1cm);
		\draw ([shift=(-120:1.5cm)]0,0) arc (-120:-60:1.5cm);
		\draw ([shift=(240:0.85cm)]0,0) arc (240:90:0.85cm);
		\draw[] (0,1.6) -- (0.9,2.1) node[pos=0,left] {$b$};
		\draw[] (0.9,2.1) -- (0.9,2.6);
		\draw (-60:1.6) -- (2.3,-2) node[pos=1,below] {$a$};
		\draw (-60:1.4) --(1.5,0) node[right]{$a$} -- (0.9,2.1);
		\node[rotate=60] at (-1,0.3) {$b+y$};
		\node at (1.15,0.3) {$y$};
		\node at (0,-1.35) {$x$};
		\end{tikzpicture}
		=\frac{1}{4}\sum_{x,y}(-1)^{(a+b)x}
		\begin{tikzpicture}[scale=.75,baseline=(current bounding box.center)]
		\draw (0,0) circle (0.5cm);
		\draw[red,line width=0.4mm]
		\foreach \a in {-60, -120} {
			(\a:0.5) -- (\a:2.5)
		};
		\draw[] (90:0.5) -- (90:1.6) node[pos=1,above]{$a+b$};
		\draw ([shift=(-60:1cm)]0,0) arc (-60:90:1cm);
		\draw ([shift=(-120:1.5cm)]0,0) arc (-120:-60:1.5cm);
		\draw ([shift=(240:0.85cm)]0,0) arc (240:90:0.85cm);
		\draw (-60:1.6) -- (2.3,-2) node[pos=1,below] {$a$};
		\node[rotate=60] at (-1,0.3) {$b+y$};
		\node[rotate=-60] at (1.15,0.3) {$y+a$};
		\node at (0,-1.35) {$x$};
		\end{tikzpicture}\\
		&=\frac{1}{4}\sum_{x,y^\prime}(-1)^{(a+b)x}
		\begin{tikzpicture}[scale=.75,baseline=(current bounding box.center)]
		\draw (0,0) circle (0.5cm);
		\draw[red,line width=0.4mm]
		\foreach \a in {-60, -120} {
			(\a:0.5) -- (\a:2.5)
		};
		\draw[] (90:0.5) -- (90:1.6) node[pos=1,above]{$a+b$};
		\draw ([shift=(-60:1cm)]0,0) arc (-60:90:1cm);
		\draw ([shift=(-120:1.5cm)]0,0) arc (-120:-60:1.5cm);
		\draw ([shift=(240:0.85cm)]0,0) arc (240:90:0.85cm);
		\draw (-60:1.6) -- (2.3,-2) node[pos=1,below] {$a$};
		\node[rotate=60] at (-1,0.3) {$a+b+y^\prime$};
		\node at (1.15,0.3) {$y^\prime$};
		\node at (0,-1.35) {$x$};
		\end{tikzpicture}
		=\begin{tikzpicture}[xscale=-1,scale=0.7,baseline=(current bounding box.center)]
		\draw[] (0,0) -- (0,1) node[pos=1,above]{$a+b$};
		\draw[red,line width=0.4mm] (0,0) -- (-0.707107,-0.707107);
		\draw[] (-0.707107,-0.707107) -- (-1.41421,-1.41421);
		\draw[red,line width=0.4mm] (-0.707107,-0.707107) -- (0,-1.41421);
		\draw[red,line width=0.4mm] (0,0) -- (0.707107,-0.707107);
		\draw[red,line width=0.4mm] (0.707107,-0.707107) -- (1.41421,-1.41421);
		\node at (-1.41421,-1.7) {$a$};
		\node at (0,-1.7) {$*$};
		\node at (1.41421,-1.7) {$*$};
		\end{tikzpicture}
	\end{align}
so it follows that 
	\begin{equation}
		\left(F_{**a}^{a+b}\right)_{b,*}=1.
	\end{equation}
The computation of $\left(F_{a**}^{a+b}\right)_{*,b}=1$ is identical. A similar computation yields $\left(F_{*a*}^{b}\right)_{*,*}=(-1)^{ab}$.

Attempting to use \eqref{eq:completeness} to compute $F_{***}^*$ is circular, so we need a different technique. There is an action of a 4-string annular category on the fusion trees on both sides of the equation
\begin{equation}
\begin{tikzpicture}[scale=0.7,baseline=(current bounding box.center)]
\draw[red,line width=0.4mm] (0,0) -- (0,1) node[pos=1,above,text=black]{$*$};
\draw[] (0,0) -- (-0.707107,-0.707107) node [pos=.5,left] {$a$};
\draw[red,line width=0.4mm] (-0.707107,-0.707107) -- (-1.41421,-1.41421);
\draw[red,line width=0.4mm] (-0.707107,-0.707107) -- (0,-1.41421);
\draw[red,line width=0.4mm] (0,0) -- (0.707107,-0.707107);
\draw[red,line width=0.4mm] (0.707107,-0.707107) -- (1.41421,-1.41421);
\node at (-1.41421,-1.7) {$*$};
\node at (0,-1.7) {$*$};
\node at (1.41421,-1.7) {$*$};
\end{tikzpicture}
=
\left(F_{***}^*\right)_{a,0}
\begin{tikzpicture}[xscale=-1,scale=0.7,baseline=(current bounding box.center)]
\draw[red,line width=0.4mm] (0,0) -- (0,1) node[pos=1,above,text=black]{$*$};
\draw[] (0,0) -- (-0.707107,-0.707107) node [pos=.5,left] {$0$};
\draw[red,line width=0.4mm] (-0.707107,-0.707107) -- (-1.41421,-1.41421);
\draw[red,line width=0.4mm] (-0.707107,-0.707107) -- (0,-1.41421);
\draw[red,line width=0.4mm] (0,0) -- (0.707107,-0.707107);
\draw[red,line width=0.4mm] (0.707107,-0.707107) -- (1.41421,-1.41421);
\node at (-1.41421,-1.7) {$*$};
\node at (0,-1.7) {$*$};
\node at (1.41421,-1.7) {$*$};
\end{tikzpicture}
+
\left(F_{***}^*\right)_{a,1}
\begin{tikzpicture}[xscale=-1,scale=0.7,baseline=(current bounding box.center)]
\draw[red] (0,0) -- (0,1) node[pos=1,above,text=black]{$*$};
\draw[] (0,0) -- (-0.707107,-0.707107) node [pos=.5,left] {$1$};
\draw[red,line width=0.4mm] (-0.707107,-0.707107) -- (-1.41421,-1.41421);
\draw[red,line width=0.4mm] (-0.707107,-0.707107) -- (0,-1.41421);
\draw[red,line width=0.4mm] (0,0) -- (0.707107,-0.707107);
\draw[red,line width=0.4mm] (0.707107,-0.707107) -- (1.41421,-1.41421);
\node at (-1.41421,-1.7) {$*$};
\node at (0,-1.7) {$*$};
\node at (1.41421,-1.7) {$*$};
\end{tikzpicture}.
\end{equation}
Both sides of the equation transform in the same representation of this 4-string action. We use this to find the $F$-symbols. Consider applying the annulus
\begin{equation}
\begin{tikzpicture}[scale=1,baseline=(current bounding box.center)]
\draw[red,line width=0.4mm](-.9,0)--(-.9,-1);
\draw[red,line width=0.4mm](0,0)--(0,-1);\draw[red,line width=0.4mm](0,.2)--(0,1);
\draw[red,line width=0.4mm](.9,0)--(.9,-1);
\draw[rounded corners] (0,.7)--(-1.6,0)--(-.9,-.5)node[pos=.5,left]{$x_0$};
\draw[rounded corners] (0,.9)--(1.6,0)--(.9,-.5)node[pos=.5,right]{$\ x_1$};
\draw[rounded corners] (-.9,-.9)--(0,-.6)node[pos=.5,above]{$x_2$};
\draw[rounded corners] (0,-.9)--(.9,-.6)node[pos=.5,above]{$x_3$};
\draw[rounded corners] (-2,-1) rectangle (2,1);
\filldraw[fill=white,rounded corners] (-1,-.2) rectangle (1,.2);
\end{tikzpicture}\label{eqn:Ftransform}
\end{equation}
to both sides, resolving the vertices where necessary. The equation becomes
\begin{align}
(-1)&^{(a+x_0)(x_1+x_2)}
\begin{tikzpicture}[scale=0.6,baseline=(current bounding box.center)]
\draw[red,line width=0.4mm] (0,0) -- (0,1) node[pos=1,above,text=black]{$*$};
\draw[] (0,0) -- (-0.707107,-0.707107) node [pos=.5,left] {$a+x_0+x_3$};
\draw[red,line width=0.4mm] (-0.707107,-0.707107) -- (-1.41421,-1.41421);
\draw[red,line width=0.4mm] (-0.707107,-0.707107) -- (0,-1.41421);
\draw[red,line width=0.4mm] (0,0) -- (0.707107,-0.707107);
\draw[red,line width=0.4mm] (0.707107,-0.707107) -- (1.41421,-1.41421);
\node at (-1.41421,-1.7) {$*$};
\node at (0,-1.7) {$*$};
\node at (1.41421,-1.7) {$*$};
\end{tikzpicture}\\[10pt]
\begin{split}
&\hspace{20pt}=
\left(F_{***}^*\right)_{a,0}(-1)^{(x_1+x_2)x_3}
\begin{tikzpicture}[xscale=-1,scale=0.6,baseline=(current bounding box.center)]
\draw[red,line width=0.4mm] (0,0) -- (0,1) node[pos=1,above,text=black]{$*$};
\draw[] (0,0) -- (-0.707107,-0.707107) node [pos=.5,right] {$x_1+x_2$};
\draw[red,line width=0.4mm] (-0.707107,-0.707107) -- (-1.41421,-1.41421);
\draw[red,line width=0.4mm] (-0.707107,-0.707107) -- (0,-1.41421);
\draw[red,line width=0.4mm] (0,0) -- (0.707107,-0.707107);
\draw[red,line width=0.4mm] (0.707107,-0.707107) -- (1.41421,-1.41421);
\node at (-1.41421,-1.7) {$*$};
\node at (0,-1.7) {$*$};
\node at (1.41421,-1.7) {$*$};
\end{tikzpicture}
\\&\hspace{40pt}+
\left(F_{***}^*\right)_{a,1}(-1)^{x_0+(1+x_1+x_2)x_3}
\begin{tikzpicture}[xscale=-1,scale=0.6,baseline=(current bounding box.center)]
\draw[red,line width=0.4mm] (0,0) -- (0,1) node[pos=1,above,text=black]{$*$};
\draw[] (0,0) -- (-0.707107,-0.707107) node [pos=.5,right] {$1+x_1+x_2$};
\draw[red,line width=0.4mm] (-0.707107,-0.707107) -- (-1.41421,-1.41421);
\draw[red,line width=0.4mm] (-0.707107,-0.707107) -- (0,-1.41421);
\draw[red,line width=0.4mm] (0,0) -- (0.707107,-0.707107);
\draw[red,line width=0.4mm] (0.707107,-0.707107) -- (1.41421,-1.41421);
\node at (-1.41421,-1.7) {$*$};
\node at (0,-1.7) {$*$};
\node at (1.41421,-1.7) {$*$};
\end{tikzpicture}
\end{split}
\\[10pt]
\begin{split}
&\hspace{20pt}=(-1)^{(a+x_0)(x_1+x_2)}
	\left(
	\left(F_{***}^*\right)_{a+x_0+x_3,x_1+x_2}
	\begin{tikzpicture}[xscale=-1,scale=0.6,baseline=(current bounding box.center)]
	\draw[red,line width=0.4mm] (0,0) -- (0,1) node[pos=1,above,text=black]{$*$};
	\draw[] (0,0) -- (-0.707107,-0.707107) node [pos=.5,right] {$x_1+x_2$};
	\draw[red,line width=0.4mm] (-0.707107,-0.707107) -- (-1.41421,-1.41421);
	\draw[red,line width=0.4mm] (-0.707107,-0.707107) -- (0,-1.41421);
	\draw[red,line width=0.4mm] (0,0) -- (0.707107,-0.707107);
	\draw[red,line width=0.4mm] (0.707107,-0.707107) -- (1.41421,-1.41421);
	\node at (-1.41421,-1.7) {$*$};
	\node at (0,-1.7) {$*$};
	\node at (1.41421,-1.7) {$*$};
	\end{tikzpicture}
	\right.
	\\&\hspace{40pt}+
	\left(F_{***}^*\right)_{a+x_0+x_3,1+x_1+x_2}\left.
	\begin{tikzpicture}[xscale=-1,scale=0.6,baseline=(current bounding box.center)]
	\draw[red,line width=0.4mm] (0,0) -- (0,1) node[pos=1,above,text=black]{$*$};
	\draw[] (0,0) -- (-0.707107,-0.707107) node [pos=.5,right] {$1+x_1+x_2$};
	\draw[red,line width=0.4mm] (-0.707107,-0.707107) -- (-1.41421,-1.41421);
	\draw[red,line width=0.4mm] (-0.707107,-0.707107) -- (0,-1.41421);
	\draw[red,line width=0.4mm] (0,0) -- (0.707107,-0.707107);
	\draw[red,line width=0.4mm] (0.707107,-0.707107) -- (1.41421,-1.41421);
	\node at (-1.41421,-1.7) {$*$};
	\node at (0,-1.7) {$*$};
	\node at (1.41421,-1.7) {$*$};
	\end{tikzpicture}\right)
\end{split}
\end{align}
where in the second step, \eqref{eqn:Ftransform} has been applied to the left-hand side. This equation reduces to
\begin{align}
\left(F_{***}^*\right)_{a,b}&=(-1)^{ab}\left(F_{***}^*\right)_{0,0}.
\end{align}
We cannot fix $\left(F_{***}^*\right)_{0,0}$ by simply asking which representation the vectors transform under, since there is a global rescaling freedom. Our normalization \eqref{eq:completeness} fixes 
\begin{align}
\left(F_{***}^*\right)_{a,b}&=(-1)^{ab}\frac{\varkappa_*}{\sqrt{2}},
\end{align}
where $\varkappa_*=\pm1$ is the \emph{Frobenius-Schur} indicator of $*$.

\subsection*{Step 6: Check if $F$-symbols obey the pentagon equation.} 
We now have the full set of $F$-symbols
	\begin{align*}
		\left(F_{abc}^{a+b+c}\right)_{a+b,b+c}&=1\\
		\left(F_{ab*}^*\right)_{a+b,*}&=1\\
		\left(F_{a*b}^*\right)_{*,*}&=(-1)^{ab}\\
		\left(F_{*ab}^*\right)_{*,a+b}&=1\\
		\left(F_{a**}^{a+b}\right)_{*,b}&=1\\
		\left(F_{*a*}^b\right)_{*,*}&=(-1)^{ab}\\
		\left(F_{**a}^{a+b}\right)_{b,*}&=1\\
		\left(F_{***}^*\right)_{a,b}&=\frac{(-1)^{ab}\varkappa_*}{\sqrt{2}},
	\end{align*}
which happen to be the $F$-symbols for the Ising category. 
\subsection{Action of the Hamiltonian}

As we have now defined how the Hamiltonian \eqref{VecZ2Hamiltonian} 
\begin{equation*}
H=-\sum \frac{1}{\sqrt{2}}
\begin{tikzpicture}[scale=0.5,baseline=(current bounding box.center)]
\draw[red,line width=0.4mm] (0,0) -- (-0.7,0.9);
\draw[red,line width=0.4mm] (0,0) -- (0.7,0.9);
\draw (0,0) to node[right] {$0$} (0,-1);
\draw[red,line width=0.4mm] (0,-1) -- (-0.7,-1.9);
\draw[red,line width=0.4mm] (0,-1) -- (0.7,-1.9);
\end{tikzpicture}
\end{equation*}
\noindent
can be realized in the framework of the $\Vec(\Z/2\Z)$ chain, we can study its action on the chain. For this purpose, we first need to change the basis using the identity
	\begin{equation}
		\begin{tikzpicture}[scale=0.5,baseline=(current bounding box.center)]
			\draw (0,0) -- (-0.7,0.9);
			\draw (0,0) -- (0.7,0.9);
			\draw (0,0) to node[right] {$e$} (0,-1);
			\draw (0,-1) -- (-0.7,-1.9);
			\draw (0,-1) -- (0.7,-1.9);
			\node at (0.8,1.2) {$b$};
			\node at (-0.8,1.2) {$a$};
			\node at (0.8,-2.2) {$d$};
			\node at (-0.8,-2.2) {$c$};
		\end{tikzpicture}=\sum_f\left(F_d^{abc}\right)_{ef}
		\begin{tikzpicture}[scale=0.5,baseline=(current bounding box.center)]
			\draw (0,0) -- (-0.9,0.7);
			\draw (0,0) -- (-0.9,-0.7);
			\draw (0,0) to node[above] {$f$} (1,0);
			\draw (1,0) -- (1.9,0.7);
			\draw (1,0) -- (1.9,-0.7);
			\node at (-1.2,0.8) {$a$};
			\node at (-1.2,-0.8) {$c$};
			\node at (2.2,0.8) {$b$};
			\node at (2.2,-0.8) {$d$};
		\end{tikzpicture}
	\end{equation}
which yields
	\begin{align}
		H&=-\sum \frac{1}{\sqrt{2}}\left(\left(F_*^{***}\right)_{00}
		\begin{tikzpicture}[scale=0.5,baseline=(current bounding box.center)]
			\draw[red,line width=0.4mm] (0,0) -- (0,2);
			\draw[red,line width=0.4mm] (1.5,0) -- (1.5,2);
		\end{tikzpicture}+\left(F_*^{***}\right)_{01}
		\begin{tikzpicture}[scale=0.5,baseline=(current bounding box.center)]
			\draw[red,line width=0.4mm] (0,0) -- (0,2);
			\draw (0,1) to node[above] {$1$} (1.5,1);
			\draw[red,line width=0.4mm] (1.5,0) -- (1.5,2);
		\end{tikzpicture}\ \right)\\
		&=-\sum\frac{1}{\sqrt{2}}\left(\ 
		\begin{tikzpicture}[scale=0.5,baseline=(current bounding box.center)]
			\draw[red,line width=0.4mm] (0,0) -- (0,2);
			\draw[red,line width=0.4mm] (1.5,0) -- (1.5,2);
		\end{tikzpicture}+
		\begin{tikzpicture}[scale=0.5,baseline=(current bounding box.center)]
			\draw[red,line width=0.4mm] (0,0) -- (0,2);
			\draw (0,1) to node[above] {$1$} (1.5,1);
			\draw[red,line width=0.4mm] (1.5,0) -- (1.5,2);
		\end{tikzpicture}\ \right),
	\end{align}
where we have used the $F$-symbols we have computed above. 

The general Hilbert space of one site of the chain is $\mathbb{C}^2\oplus \mathbb{C}$ (see \eqref{eq:configs}), hence the space for three sites is
	\begin{align}\label{eq:hspaces}
		(\mathbb{C}^2\oplus \mathbb{C})\otimes(\mathbb{C}^2\oplus \mathbb{C})\otimes(\mathbb{C}^2\oplus \mathbb{C})=(\mathbb{C}^2\otimes\mathbb{C}\otimes\mathbb{C}^2)\oplus(\mathbb{C}\otimes\mathbb{C}^2\otimes\mathbb{C})\oplus\textrm{forbidden states},
	\end{align}
where forbidden states include, for example, those where all three objects are from the bimodule.
According to the right-hand side of \eqref{eq:hspaces}, valid local configurations of the chain are of the form
	\begin{equation*}
		\begin{tikzpicture}[scale=1,baseline=(current bounding box.center)]
			\draw (-0.5,0) to node [below] {$a$} (0,0);
			\draw [red, line width=0.25mm] (0,0) -- (1,0);
			\draw (1,0) to node [below] {$b$} (1.5,0);
			\draw [red, line width=0.25mm] (0,0) -- (0,1);
			\draw [red, line width=0.25mm] (1,0) -- (1,1);
		\end{tikzpicture} \hspace{20pt}\text{and}\hspace{20pt}
		\begin{tikzpicture}[scale=1,baseline=(current bounding box.center)]
			\draw [red, line width=0.25mm] (-0.5,0) -- (0,0);
			\draw (0,0) to node [below] {$a$} (1,0);
			\draw [red, line width=0.25mm] (1,0) -- (1.5,0);
			\draw [red, line width=0.25mm] (0,0) -- (0,1);
			\draw [red, line width=0.25mm] (1,0) -- (1,1);
		\end{tikzpicture}
	\end{equation*}
\noindent
with $a,b\in\{0,1\}$. The local Hamiltonian 
	\begin{equation}
		h_i = -\frac{1}{\sqrt{2}}\left(\ 
		\begin{tikzpicture}[scale=0.5,baseline=(current bounding box.center)]
		\draw[red,line width=0.4mm] (0,0) -- (0,2);
		\draw[red,line width=0.4mm] (1.5,0) -- (1.5,2);
		\node at (0,-0.5) {$i$};
		\node at (1.5,-0.5) {$i+1$};
		\end{tikzpicture}+
		\begin{tikzpicture}[scale=0.5,baseline=(current bounding box.center)]
		\draw[red,line width=0.4mm] (0,0) -- (0,2);
		\draw (0,1) to node[above] {$1$} (1.5,1);
		\draw[red,line width=0.4mm] (1.5,0) -- (1.5,2);
		\node at (0,-0.5) {$i$};
		\node at (1.5,-0.5) {$i+1$};
		\end{tikzpicture}\ \right)
	\end{equation}
acts on these configurations in the following way: the first term of the Hamiltonian leaves the configurations invariant, while applying the second term yields
	\begin{align}\label{eq:action}
		\begin{tikzpicture}[scale=1,baseline=(current bounding box.center)]
			\draw (-0.5,0) to node [below] {$a$} (0,0);
			\draw [red, line width=0.25mm] (0,0) -- (1,0);
			\draw (1,0) to node [below] {$b$} (1.5,0);
			\draw [red, line width=0.25mm] (0,0) -- (0,1);
			\draw (0,0.5) to node[above] {$1$} (1,0.5);
			\draw [red, line width=0.25mm] (1,0) -- (1,1);
		\end{tikzpicture}&=
		\begin{tikzpicture}[scale=1,baseline=(current bounding box.center)]
			\draw (-0.5,0) to node [below] {$a$} (0,0);
			\draw [red, line width=0.25mm] (0,0) -- (1,0);
			\draw (1,0) to node [below] {$b$} (1.5,0);
			\draw [red, line width=0.25mm] (0,0) -- (0,1);
			\draw [red, line width=0.25mm] (1,0) -- (1,1);
			\draw (0,0.5) to[bend left] node[above] {$1$} (0.4,0);
			\draw (1,0.5) to[bend right] node[above] {$1$} (0.6,0);
		\end{tikzpicture}=(-1)^{a+b}
		\begin{tikzpicture}[scale=1,baseline=(current bounding box.center)]
			\draw (-0.5,0) to node [below] {$a$} (0,0);
			\draw [red, line width=0.25mm] (0,0) -- (1,0);
			\draw (1,0) to node [below] {$b$} (1.5,0);
			\draw [red, line width=0.25mm] (0,0) -- (0,1);
			\draw [red, line width=0.25mm] (1,0) -- (1,1);
		\end{tikzpicture}\equiv Z_a\otimes Z_b\\\label{eq:action2}
		\begin{tikzpicture}[scale=1,baseline=(current bounding box.center)]
			\draw [red, line width=0.25mm] (-0.5,0) -- (0,0);
			\draw (0,0) to node[below] {$a$} (1,0);
			\draw [red, line width=0.25mm] (1,0) -- (1.5,0);
			\draw [red, line width=0.25mm] (0,0) -- (0,1);
			\draw [red, line width=0.25mm] (1,0) -- (1,1);
			\draw (0,0.5) to node[above] {$1$} (1,0.5);
		\end{tikzpicture}&=
		\begin{tikzpicture}[scale=1,baseline=(current bounding box.center)]
			\draw [red, line width=0.25mm] (-0.5,0) -- (0,0);
			\draw (0,0) to node[below] {$a+1$} (1,0);
			\draw [red, line width=0.25mm] (1,0) -- (1.5,0);
			\draw [red, line width=0.25mm] (0,0) -- (0,1);
			\draw [red, line width=0.25mm] (1,0) -- (1,1);
		\end{tikzpicture}\equiv X_a
	\end{align}
A general basis state that lives in $(\mathbb{C}^2\oplus \mathbb{C})\otimes(\mathbb{C}^2\oplus \mathbb{C})\otimes(\mathbb{C}^2\oplus \mathbb{C})$ is of the form
	\begin{equation}
		\lvert x_{i-1}, x_i, x_{i+1}\rangle=\begin{tikzpicture}[scale=1,baseline=(current bounding box.center)]
		\draw [blue, line width=0.25mm] (-0.5,0) to node[below, near start] {$x_{i-1}$} (0,0);
		\draw [blue, line width=0.25mm] (0,0) to node[below] {$x_i$} (1,0);
		\draw [blue, line width=0.25mm] (1,0) to node[below, near end] {$x_{i+1}$} (1.5,0);
		\draw [red, line width=0.25mm] (0,0) -- (0,1);
		\draw [red, line width=0.25mm] (1,0) -- (1,1);
		\end{tikzpicture}
	\end{equation}
with $x_{i-1}, x_i, x_{i+1}\in\{0,1,*\}$. The goal is now to find an operator that acts on this basis that gives exactly what we have derived in \eqref{eq:action} and \eqref{eq:action2}, which means we want an expression of $h_i$ of the form
	\begin{equation}
		h_i\lvert x_{i-1},x_i,x_{i+1}\rangle=-\frac{1}{\sqrt{2}}\left(\mathbb{I}+A\right)\lvert x_{i-1},x_i,x_{i+1}\rangle
	\end{equation}
with some operator $A$. Note that we do not need to worry about how it acts on forbidden state, e.g.\ $\lvert *,*,*\rangle$. Since there are two possible configurations, we distinguish between these two cases:
	\begin{enumerate}
		\item When $x_i=*$, we are in the case of $\eqref{eq:action}$, which means the object in the middle is projected onto the state $\lvert*\rangle\langle*\rvert$. A Pauli-$Z$ operator acts on the objects $x_{i-1}$ and $x_{i+1}$, but only on the $\mathbb{C}^2$-part of the Hilbert space. Hence, we additionally need an operator that acts on $\mathbb{C}$, but since we projected the site $i$ onto $\lvert *\rangle$, and any states of the form $\lvert *,*,\cdot\rangle$ and $\lvert \cdot,*,*\rangle$ are forbidden anyway, it does not matter which operator we put here and we just denote it $\hat{O}$\footnote{We can pick $\hat{O}=0$ for simplicity.}. Following these arguments, the resulting expression for the operator $A$ is
			\begin{equation}
				A=(Z\oplus\hat{O})\otimes\lvert*\rangle\langle*\rvert\otimes(Z\oplus\hat{O}).
			\end{equation}
		\item When $x_{i-1}=x_{i+1}=*$, the outer sited are projected onto $\lvert*\rangle\langle*\rvert$ and as we have seen in \eqref{eq:action2}, the Hamiltonian acts as a Pauli-$X$ operator on the middle site. For the same reasons as above, we add the operator $\hat{O}$ and get
			\begin{equation}
				A=\lvert*\rangle\langle*\rvert\otimes(X\oplus\hat{O})\otimes\lvert*\rangle\langle*\rvert.
			\end{equation}
	\end{enumerate}
The resulting local Hamiltonian is
	\begin{equation}
		h_i=-\frac{1}{\sqrt{2}}\left(\mathbb{I}+(Z\oplus\hat{O})\otimes\lvert*\rangle\langle*\rvert\otimes(Z\oplus\hat{O})+\lvert*\rangle\langle*\rvert\otimes(X\oplus\hat{O})\otimes\lvert*\rangle\langle*\rvert\right).
	\end{equation}
If we fix some edge, e.g.\ the boundaries, to $*$ we restrict ourselves to states of the form
	
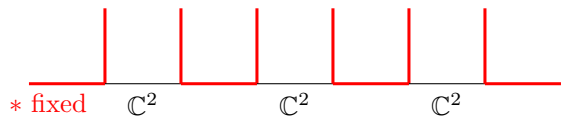
\begin{figure}[H]
		\begin{tikzpicture}[baseline=(current bounding box.center)]
		\draw[red,line width=0.4mm] (0.5,0) to node[below,near start] {$*$ fixed} (1.5,0);
		\draw[red,line width=0.4mm] (1.5,0) -- (1.5,1);
		\draw[black] (1.5,0) to node [below] {$\mathbb{C}^2$} (2.5,0);
		\draw[red,line width=0.4mm] (2.5,0) -- (2.5,1);
		\draw[red,line width=0.4mm] (2.5,0) -- (3.5,0);
		\draw[red,line width=0.4mm] (3.5,0) -- (3.5,1);
		\draw[black] (3.5,0) to node [below] {$\mathbb{C}^2$} (4.5,0);
		\draw[red,line width=0.4mm] (4.5,0) -- (4.5,1);
		\draw[red,line width=0.4mm] (4.5,0) -- (5.5,0);
		\draw[red,line width=0.4mm] (5.5,0) -- (5.5,1);
		\draw[black] (5.5,0) to node [below] {$\mathbb{C}^2$} (6.5,0);
		\draw[red,line width=0.4mm] (6.5,0) -- (6.5,1);
		\draw[red,line width=0.4mm] (6.5,0) -- (7.5,0);
		\end{tikzpicture},
	\end{figure}
\noindent
so we effectively have a system of qubits
	
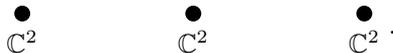
\begin{figure}[H]
		\begin{tikzpicture}[scale=1.5,baseline=(current bounding box.center)]
		\fill[black] (0,0) circle (0.07cm);
		\node at (0,-0.25) {$\mathbb{C}^2$};
		\fill[black] (1.5,0) circle (0.07cm);
		\node at (1.5,-0.25) {$\mathbb{C}^2$};
		\fill[black] (3,0) circle (0.07cm);
		\node at (3,-0.25) {$\mathbb{C}^2$};
		\end{tikzpicture}.
	\end{figure}
\noindent
On this Hilbert space, the effective Hamiltonian is 
	\begin{equation}\label{eq:eff}
		H_\mathrm{effective}=-\frac{1}{\sqrt{2}}\sum_{i\in\text{half chain}}\mathbb{I}+Z_iZ_{i+1}+X_i,
	\end{equation}
which is the Hamiltonian that corresponds to the transverse Ising model. The other possibility is to fix the boundaries to $\mathbb{C}^2$, in which case we have states of the form
	
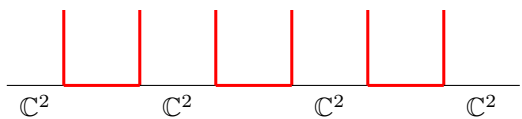
\begin{figure}[H]
		\begin{tikzpicture}
		\draw[black] (-0.25,0) to node [below] {$\mathbb{C}^2$} (0.5,0);
		\draw[red,line width=0.4mm] (0.5,0) -- (0.5,1);
		\draw[red,line width=0.4mm] (0.5,0) -- (1.5,0);
		\draw[red,line width=0.4mm] (1.5,0) -- (1.5,1);
		\draw[black] (1.5,0) to node [below] {$\mathbb{C}^2$} (2.5,0);
		\draw[red,line width=0.4mm] (2.5,0) -- (2.5,1);
		\draw[red,line width=0.4mm] (2.5,0) -- (3.5,0);
		\draw[red,line width=0.4mm] (3.5,0) -- (3.5,1);
		\draw[black] (3.5,0) to node [below] {$\mathbb{C}^2$} (4.5,0);
		\draw[red,line width=0.4mm] (4.5,0) -- (4.5,1);
		\draw[red,line width=0.4mm] (4.5,0) -- (5.5,0);
		\draw[red,line width=0.4mm] (5.5,0) -- (5.5,1);
		\draw[black] (5.5,0) to node [below] {$\mathbb{C}^2$} (6.5,0);
		\end{tikzpicture}.
	\end{figure}
\noindent
The effective Hamiltonian for these states is also of the form \eqref{eq:eff}. Hence, if we do not fix any boundaries, we get the direct sum of two copies of Ising.

\section{Conclusion}
\label{Conclusion}
In this paper, we have presented a framework to make sense of defect gauging in quantum (spin) systems. Whereas the kinematical theory can be built from distinguishable Fock space, dynamics are generally harder to implement. In order to perform this task we use a description via the ground eigenspace of a specific Hamiltonian. This reduces the problem to a combinatorial one, which can be handled by techniques from annular categories.

We have also provided the example of a one-dimensional spin chain with $\Vec(\Z/2\Z)$ fusion rules. In this case, allowing the presence of defects and fixing the boundary results in the Hamiltonian of the transverse Ising model and, hence, gives rise to a critical theory.

Since our approach is not only restricted to one-dimensional systems it would be interesting to study higher dimensional models, such as the Toric code, and add defects in this more general situation. Another interesting setting is to insert fusion categories with different fusion rules than $\Vec(\Z/2\Z)$ (e.g.\ the Fibonacci category) to our method. Even in this case, it provides a framework to build physical models with dynamically evolving defects. These theories can describe topological phases with more computational power than their counterparts without defects and are, therefore, more convenient for quantum information processing tasks.

\section*{Acknowledgments}
JCB thanks Daniel Barter for many useful discussions. This work was supported by the DFG through SFB 1227 (DQ-mat), the RTG 1991, and the cluster of excellence EXC 2123 QuantumFrontiers. Research at Perimeter Institute is supported in part by the Government of Canada through the Department of Innovation, Science and Economic Development Canada and by the Province of Ontario through the Ministry of Economic Development, Job Creation and Trade.

\bibliographystyle{halpha}
\bibliography{literature}

\newcommand{\etalchar}[1]{$^{#1}$}
\begin{thebibliography}{CGPW16}
\expandafter\ifx\csname url\endcsname\relax
  \def\url#1{\texttt{#1}}\fi
\expandafter\ifx\csname doi\endcsname\relax
  \def\doi#1{\burlalt{#1}{https://dx.doi.org/#1}}\fi
\expandafter\ifx\csname urlprefix\endcsname\relax\def\urlprefix{}\fi
\expandafter\ifx\csname href\endcsname\relax
  \def\href#1#2{#2}\fi
\expandafter\ifx\csname burlalt\endcsname\relax
  \def\burlalt#1#2{\href{#2}{#1}}\fi

\bibitem[And61]{andersonLocalizedMagneticStates1961}
P.~W. Anderson.
\newblock Localized {{Magnetic States}} in {{Metals}}.
\newblock {\em Phys. Rev.}, 124(1):41--53, 1961.

\bibitem[BASP14]{BASP14}
B.~J. Brown, A.~Al-Shimary, and J.~K. Pachos.
\newblock Entropic {Barriers} for {Two}-{Dimensional} {Quantum} {Memories}.
\newblock {\em Phys. Rev. Lett.}, 112(12):120503, 2014,
  \doi{10.1103/physrevlett.112.120503}.
\newblock \burlalt{arXiv:1307.6222}{https://arxiv.org/abs/1307.6222}.

\bibitem[BB19a]{BB19b}
J.~C. Bridgeman and D.~Barter.
\newblock {Computing data for Levin-Wen with defects}.
\newblock 2019.
\newblock \burlalt{arXiv:1907.06692}{https://arxiv.org/abs/1907.06692}.

\bibitem[BB19b]{BB19a}
J.~C. Bridgeman and D.~Barter.
\newblock {Computing Defects Associated to Bounded Domain Wall Structures: The
  $\mathrm{Vec}(\mathbb{Z}/p\mathbb{Z})$ case}.
\newblock 2019.
\newblock \burlalt{arXiv:1901.08069}{https://arxiv.org/abs/1901.08069}.

\bibitem[BBCW14]{BBCW14}
M.~Barkeshli, P.~Bonderson, M.~Cheng, and Z.~Wang.
\newblock {Symmetry} {Fractionalization}, {Defects}, and {Gauging} of
  {Topological} {Phases}.
\newblock 2014.
\newblock \burlalt{arXiv:1410.4540}{https://arxiv.org/abs/1410.4540}.

\bibitem[BBD17]{BBD17}
J.~C. Bridgeman, S.~D. Bartlett, and A.~C. Doherty.
\newblock Tensor networks with a twist: {Anyon}-permuting domain walls and
  defects in projected entangled pair states.
\newblock {\em Phys. Rev. B}, 96(24):245122, 2017,
  \doi{10.1103/physrevb.96.245122}.
\newblock \burlalt{arXiv:1708.08930}{https://arxiv.org/abs/1708.08930}.

\bibitem[BBJ]{BBJSkeletal}
D.~Barter, J.~C. Bridgeman, and C.~Jones.
\newblock In preparation.

\bibitem[BBJ18]{BBJ18}
J.~C. Bridgeman, D.~Barter, and C.~Jones.
\newblock {Fusing Binary Interface Defects in Topological Phases: The
  $\mathrm{Vec}(\mathbb{Z}/p\mathbb{Z})$ case}.
\newblock 2018.
\newblock \burlalt{arXiv:1810.09469}{https://arxiv.org/abs/1810.09469}.

\bibitem[BBJ19]{BBJ19}
D.~Barter, J.~C. Bridgeman, and C.~Jones.
\newblock Domain {W}alls in {T}opological {P}hases and the {B}rauer-{P}icard
  {R}ing for $\mathrm{Vec}(\mathbb{Z}/p\mathbb{Z})$.
\newblock {\em Commun. Math. Phys.}, 369(3):1167--1185, 2019,
  \doi{10.1007/s00220-019-03338-2}.
\newblock \burlalt{arXiv:1806.01279}{https://arxiv.org/abs/1806.01279}.

\bibitem[BC17]{bridgemanHandwavingInterpretiveDance2017}
J.~C. Bridgeman and C.~T. Chubb.
\newblock {Hand-waving and interpretive dance: an introductory course on tensor
  networks}.
\newblock {\em J. Phys. A: Math. Theor.}, 50(22):223001, 2017,
  \doi{10.1088/1751-8121/aa6dc3}.
\newblock \burlalt{arXiv:1603.03039}{https://arxiv.org/abs/1603.03039}.

\bibitem[Bea99]{Beachy1999}
J.~A. Beachy.
\newblock {\em Introductory {Lectures} on {Rings} and {Modules}}.
\newblock Cambridge University Press, 1999.

\bibitem[BJ19]{BJ19}
M.~Bischoff and C.~Jones.
\newblock Computing fusion rules for spherical {G}-extensions of fusion
  categories.
\newblock 2019.
\newblock \burlalt{arXiv:1909.02816}{https://arxiv.org/abs/1909.02816}.

\bibitem[BJQ13a]{BJQ13}
M.~Barkeshli, C.-M. Jian, and X.-L. Qi.
\newblock Genons, twist defects, and projective non-{A}belian braiding
  statistics.
\newblock {\em Phys. Rev. B}, 87(4):045130, 2013,
  \doi{10.1103/PhysRevB.87.045130}.
\newblock \burlalt{arXiv:1208.4834}{https://arxiv.org/abs/1208.4834}.

\bibitem[BJQ13b]{BJQ13b}
M.~Barkeshli, C.-M. Jian, and X.-L. Qi.
\newblock Theory of defects in {Abelian} topological states.
\newblock {\em Phys. Rev. B}, 88(23):235103, 2013,
  \doi{10.1103/physrevb.88.235103}.
\newblock \burlalt{arXiv:1305.7203}{https://arxiv.org/abs/1305.7203}.

\bibitem[BLKW17]{BLKW17}
B.~J. Brown, K.~Laubscher, M.~S. Kesselring, and J.~R. Wootton.
\newblock Poking {Holes} and {Cutting} {Corners} to {Achieve} {Clifford}
  {Gates} with the {Surface} {Code}.
\newblock {\em Phys. Rev. X}, 7(2):021029, 2017,
  \doi{10.1103/physrevx.7.021029}.
\newblock \burlalt{arXiv:1609.04673}{https://arxiv.org/abs/1609.04673}.

\bibitem[BLP{\etalchar{+}}16]{BLPSW16}
B.~J. Brown, D.~Loss, J.~K. Pachos, C.~N. Self, and J.~R. Wootton.
\newblock Quantum memories at finite temperature.
\newblock {\em Rev. Mod. Phys.}, 88(4):045005, 2016,
  \doi{10.1103/RevModPhys.88.045005}.
\newblock \burlalt{arXiv:1411.6643}{https://arxiv.org/abs/1411.6643}.

\bibitem[Bom10]{Bombin2010}
H.~Bombin.
\newblock Topological {Order} with a {Twist}: {Ising} {Anyons} from an
  {Abelian} {Model}.
\newblock {\em Phys. Rev. Lett.}, 105(3):030403, 2010,
  \doi{10.1103/physrevlett.105.030403}.
\newblock \burlalt{arXiv:1004.1838}{https://arxiv.org/abs/1004.1838}.

\bibitem[Bon07]{Bonderson}
P.~Bonderson.
\newblock {Non-Abelian Anyons and Interferometry}, 2007.
\newblock \urlprefix\url{https://thesis.library.caltech.edu/2447/2/thesis.pdf}.

\bibitem[BRSX15]{BRSX15}
Z.~Bi, A.~Rasmussen, K.~Slagle, and C.~Xu.
\newblock Classification and description of bosonic symmetry protected
  topological phases with semiclassical nonlinear sigma models.
\newblock {\em Phys. Rev. B}, 91(13):134404, 2015,
  \doi{10.1103/physrevb.91.134404}.
\newblock \burlalt{arXiv:1309.0515}{https://arxiv.org/abs/1309.0515}.

\bibitem[BS09]{BS09}
F.~A. Bais and J.~K. Slingerland.
\newblock Condensate-induced transitions between topologically ordered phases.
\newblock {\em Phys. Rev. B}, 79(4):045316, 2009,
  \doi{10.1103/physrevb.79.045316}.
\newblock \burlalt{arXiv:0808.0627}{https://arxiv.org/abs/0808.0627}.

\bibitem[BSS08]{BSS08}
P.~Bonderson, K.~Shtengel, and J.~K. Slingerland.
\newblock {Interferometry of non-Abelian anyons}.
\newblock {\em Ann. Phys.}, 323(11):2709--2755, 2008,
  \doi{10.1016/j.aop.2008.01.012}.
\newblock \burlalt{arXiv:0707.4206}{https://arxiv.org/abs/0707.4206}.

\bibitem[BSS12]{BSS12}
F.~J. Burnell, S.~H. Simon, and J.~K. Slingerland.
\newblock Phase transitions in topological lattice models via topological
  symmetry breaking.
\newblock {\em New J. Phys.}, 14(1):015004, 2012,
  \doi{10.1088/1367-2630/14/1/015004}.
\newblock \burlalt{arXiv:1012.0317}{https://arxiv.org/abs/1012.0317}.

\bibitem[BW10]{BW10}
M.~Barkeshli and X.-G. Wen.
\newblock Anyon {Condensation} and {Continuous} {Topological} {Phase}
  {Transitions} in {Non}-{Abelian} {Fractional} {Quantum} {Hall} {States}.
\newblock {\em Phys. Rev. Lett.}, 105(21):216804, 2010,
  \doi{10.1103/physrevlett.105.216804}.
\newblock \burlalt{arXiv:1007.2030}{https://arxiv.org/abs/1007.2030}.

\bibitem[BW11]{BW11}
M.~Barkeshli and X.-G. Wen.
\newblock Bilayer quantum {Hall} phase transitions and the orbifold
  non-{Abelian} fractional quantum {Hall} states.
\newblock {\em Phys. Rev. B}, 84(11):115121, 2011,
  \doi{10.1103/physrevb.84.115121}.
\newblock \burlalt{arXiv:1010.4270}{https://arxiv.org/abs/1010.4270}.

\bibitem[CCW16]{CCW16}
I.~Cong, M.~Cheng, and Z.~Wang.
\newblock Topological {Quantum} {Computation} with {Gapped} {Boundaries}.
\newblock 2016.
\newblock \burlalt{arXiv:1609.02037}{https://arxiv.org/abs/1609.02037}.

\bibitem[CCW17a]{CCW17b}
I.~Cong, M.~Cheng, and Z.~Wang.
\newblock Defects between gapped boundaries in two-dimensional topological
  phases of matter.
\newblock {\em Phys. Rev. B}, 96(19):195129, 2017,
  \doi{10.1103/physrevb.96.195129}.
\newblock \burlalt{arXiv:1703.03564}{https://arxiv.org/abs/1703.03564}.

\bibitem[CCW17b]{CCW17}
I.~Cong, M.~Cheng, and Z.~Wang.
\newblock Universal {Quantum} {Computation} with {Gapped} {Boundaries}.
\newblock {\em Phys. Rev. Lett.}, 119(17):170504, 2017,
  \doi{10.1103/physrevlett.119.170504}.
\newblock \burlalt{arXiv:1707.05490}{https://arxiv.org/abs/1707.05490}.

\bibitem[CG14]{CG14}
M.~Cheng and Z.-C. Gu.
\newblock Topological {Response} {Theory} of {Abelian} {Symmetry}-{Protected}
  {Topological} {Phases} in {Two} {Dimensions}.
\newblock {\em Phys. Rev. Lett.}, 112(14):141602, 2014,
  \doi{10.1103/physrevlett.112.141602}.
\newblock \burlalt{arXiv:1302.4803}{https://arxiv.org/abs/1302.4803}.

\bibitem[CGLW13]{CGLW13}
X.~Chen, Z.-C. Gu, Z.-X. Liu, and X.-G. Wen.
\newblock Symmetry protected topological orders and the group cohomology of
  their symmetry group.
\newblock {\em Phys. Rev. B}, 87(15):155114, 2013,
  \doi{10.1103/physrevb.87.155114}.
\newblock \burlalt{arXiv:1106.4772}{https://arxiv.org/abs/1106.4772}.

\bibitem[CGPW16]{CGPW16}
S.~X. Cui, C.~Galindo, J.~Y. Plavnik, and Z.~Wang.
\newblock On {Gauging} {Symmetry} of {Modular} {Categories}.
\newblock {\em Commun. Math. Phys.}, 348(3):1043--1064, 2016,
  \doi{10.1007/s00220-016-2633-8}.
\newblock \burlalt{arXiv:1510.03475}{https://arxiv.org/abs/1510.03475}.

\bibitem[CGW11]{CGW11}
X.~Chen, Z.-C. Gu, and X.-G. Wen.
\newblock Classification of gapped symmetric phases in one-dimensional spin
  systems.
\newblock {\em Phys. Rev. B}, 83(3):035107, 2011,
  \doi{10.1103/physrevb.83.035107}.
\newblock \burlalt{arXiv:1008.3745}{https://arxiv.org/abs/1008.3745}.

\bibitem[CSW19]{CSZW18}
S.~X. Cui, M.~{Shokrian Zini}, and Z.~Wang.
\newblock On generalized symmetries and structure of modular categories.
\newblock {\em Sci. China Math.}, 62(3):417--446, 2019,
  \doi{10.1007/s11425-018-9455-5}.
\newblock \burlalt{arXiv:1809.00245}{https://arxiv.org/abs/1809.00245}.

\bibitem[Del19]{D19}
C.~Delaney.
\newblock Fusion rules for permutation extensions of modular tensor categories.
\newblock 2019.
\newblock \burlalt{arXiv:1909.03003}{https://arxiv.org/abs/1909.03003}.

\bibitem[DIP16]{DIP16}
N.~Delfosse, P.~Iyer, and D.~Poulin.
\newblock Generalized surface codes and packing of logical qubits.
\newblock 2016.
\newblock \burlalt{arXiv:1606.07116}{https://arxiv.org/abs/1606.07116}.

\bibitem[DKLP02]{DKLP2002}
E.~Dennis, A.~Kitaev, A.~Landahl, and J.~Preskill.
\newblock Topological quantum memory.
\newblock {\em J. Math. Phys.}, 43(9):4452--4505, 2002,
  \doi{10.1063/1.1499754}.
\newblock
  \burlalt{arXiv:quant-ph/0110143}{https://arxiv.org/abs/quant-ph/0110143}.

\bibitem[DSS19]{DSPS14}
C.~L. Douglas, C.~{Schommer-Pries}, and N.~Snyder.
\newblock {The balanced tensor product of module categories}.
\newblock {\em Kyoto J. Math.}, 59(1):167--179, 2019,
  \doi{10.1215/21562261-2018-0006}.
\newblock \burlalt{arXiv:1406.4204}{https://arxiv.org/abs/1406.4204}.

\bibitem[EGNO15]{Etingof2015}
P.~Etingof, S.~Gelaki, D.~Nikshych, and V.~Ostrik.
\newblock {\em Tensor {C}ategories}, volume 205 of {\em Mathematical Surveys
  and Monographs}.
\newblock Amer. Math. Soc., 2015.
\newblock \doi{10.1090/surv/205}.

\bibitem[EH13]{EH13}
A.~M. Essin and M.~Hermele.
\newblock Classifying fractionalization: {Symmetry} classification of gapped
  $\mathbb{Z}_2$ spin liquids in two dimensions.
\newblock {\em Phys. Rev. B}, 87(10):104406, 2013,
  \doi{10.1103/physrevb.87.104406}.
\newblock \burlalt{arXiv:1212.0593}{https://arxiv.org/abs/1212.0593}.

\bibitem[EJP18]{EMJP18}
C.~{Edie-Michell}, C.~Jones, and J.~Plavnik.
\newblock {Fusion} {Rules} for $\mathbb{Z}/2\mathbb{Z}$ {Permutation}
  {Gauging}.
\newblock 2018.
\newblock \burlalt{arXiv:1804.01657}{https://arxiv.org/abs/1804.01657}.

\bibitem[EN14]{EN14}
D.~V. Else and C.~Nayak.
\newblock Classifying symmetry-protected topological phases through the
  anomalous action of the symmetry on the edge.
\newblock {\em Phys. Rev. B}, 90(23):235137, 2014,
  \doi{10.1103/physrevb.90.235137}.
\newblock \burlalt{arXiv:1409.5436}{https://arxiv.org/abs/1409.5436}.

\bibitem[ENO10]{ENO10}
P.~Etingof, D.~Nikshych, and V.~Ostrik.
\newblock Fusion categories and homotopy theory.
\newblock {\em Quantum Topol.}, 1(3):209--273, 2010, \doi{10.4171/qt/6}.
\newblock \burlalt{arXiv:0909.3140}{https://arxiv.org/abs/0909.3140}.

\bibitem[ET19]{ET19}
D.~V. Else and R.~Thorngren.
\newblock Crystalline topological phases as defect networks.
\newblock {\em Phys. Rev. B}, 99(11):115116, 2019,
  \doi{10.1103/physrevb.99.115116}.
\newblock \burlalt{arXiv:1810.10539}{https://arxiv.org/abs/1810.10539}.

\bibitem[FK10]{FK10}
L.~Fidkowski and A.~Kitaev.
\newblock Effects of interactions on the topological classification of free
  fermion systems.
\newblock {\em Phys. Rev. B}, 81(13):134509, 2010,
  \doi{10.1103/PhysRevB.81.134509}.
\newblock \burlalt{arXiv:0904.2197}{https://arxiv.org/abs/0904.2197}.

\bibitem[FK11]{FK11}
L.~Fidkowski and A.~Kitaev.
\newblock Topological phases of fermions in one dimension.
\newblock {\em Phys. Rev. B}, 83(7):075103, 2011,
  \doi{10.1103/physrevb.83.075103}.
\newblock \burlalt{arXiv:1008.4138}{https://arxiv.org/abs/1008.4138}.

\bibitem[FKLW02]{FKLW02}
M.~H. Freedman, A.~Kitaev, M.~J. Larsen, and Z.~Wang.
\newblock Topological quantum computation.
\newblock {\em Bull. Am. Math. Soc.}, 40(01):31--39, 2002,
  \doi{10.1090/s0273-0979-02-00964-3}.
\newblock
  \burlalt{arXiv:quant-ph/0101025}{https://arxiv.org/abs/quant-ph/0101025}.

\bibitem[FLW02a]{FLW02}
M.~H. Freedman, M.~Larsen, and Z.~Wang.
\newblock A {Modular} {Functor} {Which} is {Universal} for {Quantum}
  {Computation}.
\newblock {\em Commun. Math. Phys.}, 227(3):605--622, 2002,
  \doi{10.1007/s002200200645}.
\newblock
  \burlalt{arXiv:quant-ph/0001108}{https://arxiv.org/abs/quant-ph/0001108}.

\bibitem[FLW02b]{FLW02b}
M.~H. Freedman, M.~J. Larsen, and Z.~Wang.
\newblock The {Two}-{Eigenvalue} {Problem} and {Density} of {Jones}
  {Representation} of {Braid} {Groups}.
\newblock {\em Commun. Math. Phys.}, 228(1):177--199, 2002,
  \doi{10.1007/s002200200636}.
\newblock \burlalt{arXiv:math/0103200}{https://arxiv.org/abs/math/0103200}.

\bibitem[FM13]{FM13}
D.~S. Freed and G.~W. Moore.
\newblock Twisted {Equivariant} {Matter}.
\newblock {\em Ann. Henri Poincar{\'{e}}}, 14(8):1927--2023, 2013,
  \doi{10.1007/s00023-013-0236-x}.
\newblock \burlalt{arXiv:1208.5055}{https://arxiv.org/abs/1208.5055}.

\bibitem[FPSV15]{JPSV15}
J.~Fuchs, J.~Priel, C.~Schweigert, and A.~Valentino.
\newblock On the {Brauer} {Groups} of {Symmetries} of {Abelian}
  {Dijkgraaf}-{Witten} {Theories}.
\newblock {\em Commun. Math. Phys.}, 339(2):385--405, 2015,
  \doi{10.1007/s00220-015-2420-y}.
\newblock \burlalt{arXiv:1404.6646}{https://arxiv.org/abs/1404.6646}.

\bibitem[Fre98]{Freedman1998}
M.~H. Freedman.
\newblock \textit{{P}/{NP}}, and the quantum field computer.
\newblock {\em Proc. Natl. Acad. Sci. USA}, 95(1):98--101, 1998,
  \doi{10.1073/pnas.95.1.98}.

\bibitem[Fre14]{F14}
D.~S. Freed.
\newblock Short-range entanglement and invertible field theories.
\newblock 2014.
\newblock \burlalt{arXiv:1406.7278}{https://arxiv.org/abs/1406.7278}.

\bibitem[FSV13]{FSV13}
J.~Fuchs, C.~Schweigert, and A.~Valentino.
\newblock Bicategories for {Boundary} {Conditions} and for {Surface} {Defects}
  in 3-d {TFT}.
\newblock {\em Commun. Math. Phys.}, 321(2):543--575, 2013,
  \doi{10.1007/s00220-013-1723-0}.
\newblock \burlalt{arXiv:1203.4568}{https://arxiv.org/abs/1203.4568}.

\bibitem[FTL{\etalchar{+}}07]{Feiguin2007}
A.~Feiguin, S.~Trebst, A.~W.~W. Ludwig, M.~Troyer, A.~Kitaev, Z.~Wang, and
  M.~H. Freedman.
\newblock {I}nteracting {A}nyons in {T}opological {Q}uantum {L}iquids: {T}he
  {G}olden {C}hain.
\newblock {\em Phys. Rev. Lett.}, 98(16):160409, 2007,
  \doi{10.1103/physrevlett.98.160409}.
\newblock
  \burlalt{arXiv:cond-mat/0612341}{https://arxiv.org/abs/cond-mat/0612341}.

\bibitem[Hew97]{hewsonKondoProblemHeavy1997}
A.~C. Hewson.
\newblock {\em The {{Kondo Problem}} to {{Heavy Fermions}}}.
\newblock Camb. Univ. Press, 1997.

\bibitem[HK10]{HK10}
M.~Z. Hasan and C.~L. Kane.
\newblock \textit{Colloquium}: {Topological} insulators.
\newblock {\em Rev. Mod. Phys.}, 82(4):3045--3067, 2010,
  \doi{10.1103/revmodphys.82.3045}.
\newblock \burlalt{arXiv:1002.3895}{https://arxiv.org/abs/1002.3895}.

\bibitem[HW12]{HW12}
L.-Y. Hung and X.-G. Wen.
\newblock Quantized topological terms in weakly coupled gauge theories and
  their connection to symmetry protected topological phases.
\newblock 2012.
\newblock \burlalt{arXiv:1211.2767}{https://arxiv.org/abs/1211.2767}.

\bibitem[Kap14]{K14}
A.~Kapustin.
\newblock Symmetry {Protected} {Topological} {Phases}, {Anomalies}, and
  {Cobordisms}: {Beyond} {Group} {Cohomology}.
\newblock 2014.
\newblock \burlalt{arXiv:1403.1467}{https://arxiv.org/abs/1403.1467}.

\bibitem[Kit03]{Kit03}
A.~Kitaev.
\newblock Fault-tolerant quantum computation by anyons.
\newblock {\em Ann. Phys.}, 303(1):2--30, 2003,
  \doi{10.1016/S0003-4916(02)00018-0}.
\newblock
  \burlalt{arXiv:quant-ph/9707021}{https://arxiv.org/abs/quant-ph/9707021}.

\bibitem[Kit09]{Kitaev2009}
A.~Kitaev.
\newblock Periodic table for topological insulators and superconductors.
\newblock In V.~Lebedev and M.~Feigel'man, editors, {\em {AIP} Conference
  Proceedings}, volume 1134, pages 22--30, Chernogolokova, 2009. {AIP},
  \burlalt{arXiv:0901.2686}{https://arxiv.org/abs/0901.2686}.
\newblock \doi{10.1063/1.3149495}.

\bibitem[KK12]{KK12}
A.~Kitaev and L.~Kong.
\newblock Models for {Gapped} {Boundaries} and {Domain} {Walls}.
\newblock {\em Commun. Math. Phys.}, 313(2):351--373, 2012,
  \doi{10.1007/s00220-012-1500-5}.
\newblock \burlalt{arXiv:1104.5047}{https://arxiv.org/abs/1104.5047}.

\bibitem[KPEB18]{KPEB18}
M.~S. Kesselring, F.~Pastawski, J.~Eisert, and B.~J. Brown.
\newblock The boundaries and twist defects of the color code and their
  applications to topological quantum computation.
\newblock {\em Quantum}, 2:101, 2018, \doi{10.22331/q-2018-10-19-101}.
\newblock \burlalt{arXiv:1806.02820}{https://arxiv.org/abs/1806.02820}.

\bibitem[LG12]{LG12}
M.~Levin and Z.-C. Gu.
\newblock Braiding statistics approach to symmetry-protected topological
  phases.
\newblock {\em Phys. Rev. B}, 86(11):115109, 2012,
  \doi{10.1103/physrevb.86.115109}.
\newblock \burlalt{arXiv:1202.3120}{https://arxiv.org/abs/1202.3120}.

\bibitem[LS12]{LS12}
M.~Levin and A.~Stern.
\newblock Classification and analysis of two-dimensional {A}belian fractional
  topological insulators.
\newblock {\em Phys. Rev. B}, 86(11):115131, 2012,
  \doi{10.1103/physrevb.86.115131}.
\newblock \burlalt{arXiv:1205.1244}{https://arxiv.org/abs/1205.1244}.

\bibitem[LV12]{LV12}
Y.-M. Lu and A.~Vishwanath.
\newblock Theory and classification of interacting integer topological phases
  in two dimensions: {A} {C}hern-{S}imons approach.
\newblock {\em Phys. Rev. B}, 86(12):125119, 2012,
  \doi{10.1103/physrevb.86.125119}.
\newblock \burlalt{arXiv:1205.3156}{https://arxiv.org/abs/1205.3156}.

\bibitem[LV16]{LV16}
Y.-M. Lu and A.~Vishwanath.
\newblock Classification and properties of symmetry-enriched topological
  phases: {C}hern-{S}imons approach with applications to ${Z}_2$ spin liquids.
\newblock {\em Phys. Rev. B}, 93(15):155121, 2016,
  \doi{10.1103/physrevb.93.155121}.
\newblock \burlalt{arXiv:1302.2634}{https://arxiv.org/abs/1302.2634}.

\bibitem[MFCV14]{MFCV15}
M.~A. Metlitski, L.~Fidkowski, X.~Chen, and A.~Vishwanath.
\newblock Interaction effects on {3D} topological superconductors: surface
  topological order from vortex condensation, the 16 fold way and fermionic
  {Kramers} doublets.
\newblock 2014.
\newblock \burlalt{arXiv:1406.3032}{https://arxiv.org/abs/1406.3032}.

\bibitem[MR13]{MR13}
A.~Mesaros and Y.~Ran.
\newblock Classification of symmetry enriched topological phases with exactly
  solvable models.
\newblock {\em Phys. Rev. B}, 87(15):155115, 2013,
  \doi{10.1103/physrevb.87.155115}.
\newblock \burlalt{arXiv:1212.0835}{https://arxiv.org/abs/1212.0835}.

\bibitem[NCMT14]{NCMT14}
T.~Neupert, C.~Chamon, C.~Mudry, and R.~Thomale.
\newblock Wire deconstructionism of two-dimensional topological phases.
\newblock {\em Phys. Rev. B}, 90(20):205101, 2014,
  \doi{10.1103/physrevb.90.205101}.
\newblock \burlalt{arXiv:1403.0953}{https://arxiv.org/abs/1403.0953}.

\bibitem[NSS{\etalchar{+}}08]{NSSFD08}
C.~Nayak, S.~H. Simon, A.~Stern, M.~Freedman, and S.~{Das Sarma}.
\newblock {Non}-{Abelian} anyons and topological quantum computation.
\newblock {\em Rev. Mod. Phys.}, 80(3):1083--1159, 2008,
  \doi{10.1103/RevModPhys.80.1083}.
\newblock \burlalt{arXiv:0707.1889}{https://arxiv.org/abs/0707.1889}.

\bibitem[Ocn94]{ocneanu}
A.~Ocneanu.
\newblock Chirality for operator algebras.
\newblock In H.~Araki, Y.~Kawahigashi, and H.~Kosaki, editors, {\em Subfactors
  Taniguchi Symposium}, pages 39--63, Kyuzeso, 1994. World Sci. Publ.

\bibitem[Osb16]{OsborneVideoLectureAQT}
Tobias~J. Osborne.
\newblock Lecture on {Advanced} {Quantum} {Theory}, 2016.
\newblock
  \urlprefix\url{https://www.youtube.com/playlist?list=PLDfPUNusx1Eo60qx3Od2KLUL4b7VDPo9F}.

\bibitem[PY15]{PY15}
F.~Pastawski and B.~Yoshida.
\newblock Fault-tolerant logical gates in quantum error-correcting codes.
\newblock {\em Phys. Rev. A}, 91(1):012305, 2015,
  \doi{10.1103/physreva.91.012305}.
\newblock \burlalt{arXiv:1408.1720}{https://arxiv.org/abs/1408.1720}.

\bibitem[QZ11]{QZ11}
X.-L. Qi and S.-C. Zhang.
\newblock Topological insulators and superconductors.
\newblock {\em Rev. Mod. Phys.}, 83(4):1057--1110, 2011,
  \doi{10.1103/revmodphys.83.1057}.
\newblock \burlalt{arXiv:1008.2026}{https://arxiv.org/abs/1008.2026}.

\bibitem[RH07]{RH07}
R.~Raussendorf and J.~Harrington.
\newblock Fault-{Tolerant} {Quantum} {Computation} with {High} {Threshold} in
  {Two} {Dimensions}.
\newblock {\em Phys. Rev. Lett.}, 98(19):190504, 2007,
  \doi{10.1103/physrevlett.98.190504}.
\newblock
  \burlalt{arXiv:quant-ph/0610082}{https://arxiv.org/abs/quant-ph/0610082}.

\bibitem[SRFL08]{SRFL08}
A.~P. Schnyder, S.~Ryu, A.~Furusaki, and A.~W.~W. Ludwig.
\newblock Classification of topological insulators and superconductors in three
  spatial dimensions.
\newblock {\em Phys. Rev. B}, 78(19):195125, 2008,
  \doi{10.1103/physrevb.78.195125}.
\newblock \burlalt{arXiv:0803.2786}{https://arxiv.org/abs/0803.2786}.

\bibitem[Swi14]{Swingle2014}
B.~Swingle.
\newblock Interplay between short- and long-range entanglement in
  symmetry-protected phases.
\newblock {\em Phys. Rev. B}, 90(3):035451, 2014,
  \doi{10.1103/physrevb.90.035451}.
\newblock \burlalt{arXiv:1209.0776}{https://arxiv.org/abs/1209.0776}.

\bibitem[Ter15]{Ter15}
B.~M. Terhal.
\newblock {Quantum} error correction for quantum memories.
\newblock {\em Rev. Mod. Phys.}, 87(2):307--346, 2015,
  \doi{10.1103/RevModPhys.87.307}.
\newblock \burlalt{arXiv:1302.3428}{https://arxiv.org/abs/1302.3428}.

\bibitem[TPB11]{TPB11}
A.~M. Turner, F.~Pollmann, and E.~Berg.
\newblock Topological phases of one-dimensional fermions: {A}n entanglement
  point of view.
\newblock {\em Phys. Rev. B}, 83(7):075102, 2011,
  \doi{10.1103/physrevb.83.075102}.
\newblock \burlalt{arXiv:1008.4346}{https://arxiv.org/abs/1008.4346}.

\bibitem[Tur00]{T00}
V.~Turaev.
\newblock Homotopy field theory in dimension 3 and crossed group-categories.
\newblock 2000.
\newblock \burlalt{arXiv:math/0005291}{https://arxiv.org/abs/math/0005291}.

\bibitem[Tur10]{Turaev2010}
V.~Turaev.
\newblock {\em Homotopy {Quantum} {Field} {Theory}}, volume~10 of {\em EMS
  Tracts in Mathematics}.
\newblock Euro. Math. Soc., 2010.

\bibitem[TY98]{TY}
D.~Tambara and S.~Yamagami.
\newblock Tensor {Categories} with {Fusion} {Rules} of {Self}-{Duality} for
  {Finite} {Abelian} {Groups}.
\newblock {\em J. Algebra}, 209(2):692--707, 1998,
  \doi{10.1006/jabr.1998.7558}.

\bibitem[WBV17]{WBV17}
D.~J. Williamson, N.~Bultinck, and F.~Verstraete.
\newblock Symmetry-enriched topological order in tensor networks: {Defects},
  gauging and anyon condensation.
\newblock 2017.
\newblock \burlalt{arXiv:1711.07982}{https://arxiv.org/abs/1711.07982}.

\bibitem[Wen90]{Wen90}
X.-G. Wen.
\newblock {Topological} {Orders} in {Rigid} {States}.
\newblock {\em Int. J. Mod. Phys. B}, 04(02):239--271, 1990,
  \doi{10.1142/s0217979290000139}.

\bibitem[Wen02]{Wen2002}
X.-G. Wen.
\newblock Quantum orders and symmetric spin liquids.
\newblock {\em Phys. Rev. B}, 65(16):165113, 2002,
  \doi{10.1103/physrevb.65.165113}.
\newblock
  \burlalt{arXiv:cond-mat/0107071}{https://arxiv.org/abs/cond-mat/0107071}.

\bibitem[Wen07]{Wen07}
X.-G. Wen.
\newblock {\em Quantum {Field} {Theory} of {Many}-{Body} {Systems}}.
\newblock Oxford University Press, 2007.

\bibitem[Whi92]{whiteDensityMatrixFormulation1992}
S.~R. White.
\newblock Density matrix formulation for quantum renormalization groups.
\newblock {\em Phys. Rev. Lett.}, 69(10):2863--2866, 1992,
  \doi{10.1103/PhysRevLett.69.2863}.

\bibitem[Wil75]{wilsonRenormalizationGroupCritical1975}
K.~G. Wilson.
\newblock The renormalization group: {Critical} phenomena and the {Kondo}
  problem.
\newblock {\em Rev. Mod. Phys.}, 47(4):773--840, 1975,
  \doi{10.1103/RevModPhys.47.773}.

\bibitem[WN90]{WN90}
X.-G. Wen and Q.~Niu.
\newblock Ground-state degeneracy of the fractional quantum {H}all states in
  the presence of a random potential and on high-genus {R}iemann surfaces.
\newblock {\em Phys. Rev. B}, 41(13):9377--9396, 1990,
  \doi{10.1103/physrevb.41.9377}.

\bibitem[WPS14]{WPS14}
C.~Wang, A.~C. Potter, and T.~Senthil.
\newblock Classification of {Interacting} {Electronic} {Topological}
  {Insulators} in {Three} {Dimensions}.
\newblock {\em Science}, 343(6171):629--631, 2014,
  \doi{10.1126/science.1243326}.
\newblock \burlalt{arXiv:1306.3238}{https://arxiv.org/abs/1306.3238}.

\bibitem[Yos15]{Yoshida2015}
B.~Yoshida.
\newblock Topological color code and symmetry-protected topological phases.
\newblock {\em Phys. Rev. B}, 91(24):245131, 2015,
  \doi{10.1103/physrevb.91.245131}.
\newblock \burlalt{arXiv:1503.07208}{https://arxiv.org/abs/1503.07208}.

\bibitem[Yos17]{Yoshida2017}
B.~Yoshida.
\newblock Gapped boundaries, group cohomology and fault-tolerant logical gates.
\newblock {\em Ann. Phys.}, 377(2):387--413, 2017,
  \doi{10.1016/j.aop.2016.12.014}.
\newblock \burlalt{arXiv:1509.03626}{https://arxiv.org/abs/1509.03626}.

\end{thebibliography}
\end{document}